# Investigating Disagreement in the Scientific Literature

[1]*Wout S. Lamers* (0000-0001-7176-9579), [2]*Kevin Boyack* (0000-0001-7814-8951), [3]*Vincent Larivière* (0000-0002-2733-0689), [4]*Cassidy R. Sugimoto* (0000-0001-8608-3203), [1]*Nees Jan van Eck* (0000-0001-8448-4521), [1]*Ludo Waltman* (0000-0001-8249-1752), [5]*Dakota Murray* (0000-0002-7119-0169)*

[1]*Centre for Science and Technology Studies, Leiden University, Leiden, Netherlands*
[2]*SciTech Strategies, Inc., Albuquerque, NM, USA*
[3]*École de bibliothéconomie et des sciences de l'information, Université de Montréal, Canada*
[4]*School of Public Policy, Georgia Institute of Technology, Atlanta, GA, USA*
[5]*School of Informatics, Computing, and Engineering, Indiana University Bloomington, IN, USA*

*Corresponding author, dakmurra@iu.edu*

# Abstract

Disagreement is essential to scientific progress. However, the extent of disagreement in science, its evolution over time, and the fields in which it happens, remains poorly understood. Leveraging a massive collection of English-language scientific texts, we develop a cue-phrase based approach to identify instances of disagreement citations across more than four million scientific articles. Using this method, we construct an indicator of disagreement across scientific fields over the 2000-2015 period. In contrast with black-box text classification methods, our framework is transparent and easily interpretable. We reveal a disciplinary spectrum of disagreement, with higher disagreement in the social sciences and lower disagreement in physics and mathematics. However, detailed disciplinary analysis demonstrates heterogeneity across sub-fields, revealing the importance of local disciplinary cultures and epistemic characteristics of disagreement. Paper-level analysis reveals notable episodes of disagreement in science, and illustrates how methodological artifacts can confound analyses of scientific texts. These findings contribute to a broader understanding of disagreement and establish a foundation for future research to understanding key processes underlying scientific progress.

# Introduction

Disagreement is a common phenomenon in science. The history of science is ripe with histories of famous discoveries, which are often embroiled in debates, controversies, and disputes. Dialectic discourse emerged in ancient Greece, whereby the truth was thought to emerge from the arguments and counterarguments of scholars engaged in dialogue. The modern scientific method arose from a similar dialogue 350 years ago, as two individuals—Robert Boyle and Thomas Hobbes—debated over the meaning of experimental results obtained with the newly-invented air pump (Shapin & Schaffer, 2011). Disagreement anchors much of the lore surrounding major scientific discoveries. For example, Alfred Wegener's theory of plate tectonics was initially rejected by the scientific community, and physics endured a decades-long dispute over the existence of gravitational waves (Collins, 2017) and the value of the Hubble constant (Castelvecchi, 2020). Other conflicts are influenced by forces external to science, such as the controversies on the link between cigarette and lung cancer or between greenhouse gas and climate change (Oreskes & Conway, 2011). Disagreement features prominently in key theories of science, such as Popper's falsifiability (Popper, 1963), Kuhn's paradigm shifts (Kuhn, 1996), and Kitcher's scientific division of labor (Kitcher, 1995).

Despite its importance to science, however, there is little empirical evidence of how much disagreement exists, where it is most common, and its consequences. Quantitative measures can be valuable tools to better understand the role and extent of disagreement across fields of science. Previous research has focused on consensus as evidenced by citation networks (Bruggeman et al., 2012; Shwed & Bearman, 2010, 2012); on concepts related to disagreement in scientific texts such as negative citations, disputing citations, and uncertainty (Catalini et al., 2015; Chen et al., 2018; Nicholson et al., 2021); and on word-count based approaches (Bertin et al., 2016). Studying disagreement is challenging, given the lack of a widely accepted theoretical framework for conceptualizing disagreement combined with major challenges in its operationalization, for instance, the limited availability of large-scale collections of scientific texts.

This paper proposes an operationalization of disagreement in scientific articles that captures direct disagreement between two papers, as well as statements indicative of disagreement within the community. We describe a methodological approach to generate and manually-validate cue-phrases that reliably match to citation sentences (citances) to represent valid instances of disagreement. We then use this approach to quantify the extent of disagreement across more than four million publications in the Elsevier *ScienceDirect* database, and investigate the rate of disagreement across fields of science.

# Literature Review

It is widely acknowledged that disagreement plays a fundamental role in scientific progress (Balietti et al., 2015; Sarewitz, 2011; "The Power of Disagreement," 2016). However, few studies have tried to quantify the level of disagreement in the scientific literature. Part of this may be explained by the fact that disagreement is difficult to both define and measure. There have been, however, attempts to assess consensus or uncertainty in the literature. Much of the early work on consensus attempted at characterizing differences between so-called hard and soft sciences. Cole (1983) described a series of experiments done in several fields, finding no evidence of differences in cognitive consensus along the hierarchy of sciences (Comte, 1856). Hargens (1988) claimed that fields having journals with higher rejection rates had lower consensus. This claim was contested by Cole and colleagues (1988), who argued that other variables accounted for the differences, and that reviewer's assessments would be a better measure of consensus than rejection rates. Fanelli (2010) found that positive results—support for the paper's hypotheses—was far higher in the social sciences than the physical sciences, which is argued to reflect higher ambiguity, and thus lower consensus, in the social sciences.

Recent studies on scientific consensus have made use of citations and text. Through a series of case studies, Shwed and Bearman (2010) used network modularity to claim that divisions in the citation network decreased over time, corresponding to increased consensus. Nicolaisen and Frandsen (2012) used a Gini index calculated over bibliographic coupling count distributions to approximate consensus, and found that physics papers showed more consensus on average than psychology papers. Using a corpus of nearly 168,000 papers, Evans et al. (2016) calculated the Shannon entropy of language in a set of eight fields, and found more evidence that consensus was higher in the hard sciences than the social sciences.

Other studies developed methods to identify *uncertainty* in text, a concept that is related to disagreement, and a potential indicator of consensus. For example, Szarvas et al. (2012) interpreted uncertainty as a 'lack of information' and created a cue-word based uncertainty detection model based on three annotated datasets (BioScope, WikiWeasel, and Fact Bank). Their results suggest that while domain-specific cues are useful, there remain cues that can reasonably identify uncertainty across domains. Similarly, Yang et al. (2012) developed a classifier based on manually annotated uncertainty cues and conditional random fields, and conducted a series of experiments to assess the performance of their method. Chen, Song and Heo (2018) later extended these approaches and applied them to an empirical study of uncertainty across science. They first introduced a conceptual framework to study uncertainty that incorporates epistemic status and perturbation strength, and then measured uncertainty in 24 high-level scientific fields, and finally created an expanded set of uncertainty cues to support further analysis (Note that these rates included all types of uncertainty, whether they be theoretical, conceptual, or experimental, and within or between studies). The reported rate of uncertainty closely mirrored consensus, highest in the social sciences, followed by the medical sciences, environmental sciences, and engineering.

Many of the cues used as a starting point by Chen et al. (2018) are hedging terms, which are commonly used in scientific writing to express possibility rather than certainty (Hyland, 1998). In addition to being field-dependent, hedging rates have also been found to depend on whether a paper is primarily methodological. Recent work by Small and colleagues (Small, 2018; Small et al., 2019) showed that citing sentences (i.e., citances) with the word 'may' occur more frequently when citing method papers than non-method papers. More recently, Bornmann, Wray and Haunschild (2020) used a similar method to investigate uncertainty associated with specific concepts in the context of highly cited works. While some might equate uncertainty or hedging with disagreement, they are not the same. As mentioned by Small, when citing another work, "hedging does not assert that the paper is wrong, but only suggests that uncertainty surrounds some aspect of the ideas put forward" (Small et al., 2019, p. 8). Here, we attempt to explicitly identify and measure scientific disagreement by using a large set of citances across all fields and by developing a set of cues validated by expert assessment.

Other studies of disagreement have been performed in the context of classification schemes of citation function. In an early attempt to categorize types of citations, disagreement was captured as "juxtapositional" and "negational" citations (Moravcsik & Murugesan, 1975). However, this scheme was manually developed using a limited sample of papers and citations, and so the robustness and validity of the categories cannot be easily assessed. More recently, scholars have used larger datasets and machine learning techniques to scale citation classifications, often including categories of citations similar or inclusive of disagreement. For example, Teufel et al., (2006) developed a four-category scheme in which disagreement might be captured under their "weakness" or "contrast" citation types. Bertin and colleagues (2016) used n-grams to study location of negative and positive citations, and showed that that the word "*disagree**" was much less likely to occur than the world "*agree**", irrespective of papers' sections. In another study that aimed to identify meaningful citations, Valenzuela et al. (2015) captured disagreement under the "comparison" citation type. Others have sought more coarse categories: Catalini et al. (2015) classified over 750,000 references made by papers published in the *Journal of Immunology* as either positive or negative, finding that negative references comprised about two percent of all references made. However, while these machine learning approaches are useful for analyzing large text data, they are also black boxes which can obfuscate issues and limit interpretation of their results.

Building on these studies, we propose a novel approach for the study of disagreement based on a set of manually-validated cue-phrases. We conduct one of the first empirical investigations into the specific notion of *disagreement* in science, and our inclusive definition allows us to capture explicit disagreement between specific papers as well as traces of disagreement within a field. Our cue-phrase based approach is more transparent and reproducible than black-box machine learning methodologies commonly employed in citation classification, and also extensively validated using over 3,000 citation sentences representing a range of fields. We extend the scale of past analyses, identifying instances of disagreement across more than four million scientific publications.

# Materials and Methods

## Data

We sourced data from an Elsevier ScienceDirect corpus that was also used in a previous study (Boyack et al., 2018) and that is hosted at the Centre for Science and Technology Studies (CWTS) at Leiden University. This corpus contains the full-text information of nearly five million English-language research articles, short communications, and review articles published in Elsevier journals between 1980 and 2016. The corpus comprises articles from nearly 3,000 Elsevier journals. Given that Elsevier is the largest publisher in the world, this corpus is one of the largest multidisciplinary sources of full-text scientific articles currently available, with coverage of both natural sciences, medical sciences, as well as the social sciences and humanities.

We focus our analysis on sentences containing in-text citations (citances). These citances were extracted from the full-text of articles following the procedure outlined in previous work (Boyack et al., 2018). The Elsevier ScienceDirect corpus that was used was constructed in the following way. First, the Crossref REST API was used to identify all articles published by Elsevier. The full-text of these articles was subsequently downloaded from the Elsevier ScienceDirect API (Article Retrieval API) in XML format. Each XML full-text record was parsed to identify major sections and paragraphs (using XML tags), and sentences (using a sentence-splitting algorithm). In-text citations in the main text were identified by parsing the main text (excluding those in footnotes and figure and table captions). XML records without in-text citations were discarded, and publications from before 1998 were omitted from analysis due to poor availability of full-text records before that year. The resulting dataset consisted of 4,776,340 publications containing a total of 145,351,937 citances, ranging from 1998 to 2016.

To facilitate analysis at the level of scientific fields, articles in Elsevier ScienceDirect as well as the references cited in these articles were matched with records in the Web of Science[1] database based on their DOI (where available) and a combination of publication year, volume number, and first page number. We used an existing classification of research articles and review articles in the Web of Science created at CWTS. In this hierarchical classification, each article published between 2000 and 2015 and indexed in the Web of Science was algorithmically assigned to a single micro-level scientific field, each of which are in turn members of one of 817 meso-level fields. It is at this meso-level that we perform our most detailed analyses, the categories being fine-grained enough to provide insights into local communities while also large enough to contain a sufficient number of citances. A further benefit of this approach to clustering is that each meso-level field, and each individual publication, can be directly grouped into one of five broad fields: *Biomedical and Health Sciences*, *Life and Earth Sciences*, *Mathematics and Computer Science*, *Physical Sciences and Engineering*, and *Social Sciences and Humanities*. Linking our dataset to this classification system resulted in a subset of 3,883,563 papers containing 118,012,368 citances, spanning 2000 to 2015. The classification was created algorithmically based on direct citation links between articles, using the methodology introduced by Waltman and Van Eck (2012) and Traag et al. (2019). A visualization of the meso-level classification was created using the VOSviewer software (van Eck & Waltman, 2010).

[1] The Web of Science database used by CWTS includes the Science Citation Index Expanded, the Social Sciences Citation Index, and the Arts & Humanities Citation Index. Other Web of Science citation indices were not included.

## Operationalizing disagreement

Researchers can disagree for many reasons, sometimes over data and methodologies, but more often because of differences in interpretation (Dieckmann & Johnson, 2019). Some of these disagreements are explicitly hostile and adversarial, whereas others are more subtle, such as contrasting findings with past results and theories. We introduce an inclusive definition of disagreement that captures explicit textual instances of disagreement, controversy, dissonance, or lack of consensus between scientific publications, including cases where citing authors are not taking an explicit stance themselves. Our definition distinguishes between two kinds of disagreement, which together capture the diversity of obvious and subtle disagreement in the scientific literature: *paper-level disagreement* and *community-level disagreement*.

The first, *paper-level disagreement*, occurs when one publication offers a finding or perspective that is (at least partly) incompatible with the perspective of another (even though there may be no explicit contradiction). Consider the following example of a citation sentence explicitly disagreeing with the conclusion of a past study:

> *We find that coffee does not cause cancer, contrary to the finding of <ref> that coffee does cause cancer.*

*Paper-level disagreement* can also be more subtle. For example, in the following two disagreement sentences, although they do not resolutely contradict one another, the citing and cited publications use models that are based on incompatible assumptions (first sentence), or observe different effects from different data (second sentence):

> *Assuming that coffee increases the probability of cancer by 50%, the predicted life expectancy for the Dutch population is 80 years, in contrast to the 85 years proposed by models that assumed coffee does not increase the risk of cancer <ref>.*

> *Contrary to previous studies that did not observe evidence to support the hypothesis that coffee causes cancer <ref>, our data suggests that drinking coffee increases the probability of cancer by 50%.*

*Community-level disagreement,* in contrast, refers to the situation in which a citing publication, without explicitly disagreeing with a cited publication, instead draws attention to a controversy or lack of consensus in the larger body of literature. Including community-level disagreement allows us to identify indirect traces of disagreement in a field, even in the absence of explicit disagreement between the referenced authors, or between the citing and cited papers. Consider the following examples of community-level disagreement; the first notes the disagreement between the referenced studies, and the second cites a single review article indicating disagreement within the field:

> *There remains controversy in the scientific literature over whether or not coffee is associated with an increased risk of cancer <refs>.*

> *A recent review of studies assessing the potential link between coffee consumption and cancer risk has observed continued controversy <ref>.*

Here, we do not differentiate between paper-level or community-level disagreement, including both under our operationalization of disagreement.

## Signal and filter terms

We compose cue-phrases of *signal* terms and *filter* terms. A variety of approaches can be used to generate these terms, and our approach is not dependent on any particular strategy. Here, we create a preliminary set of signal terms through an intensive iterative process of manually identifying, classifying, validating, and deliberating on strategies for identifying instances of disagreement. This took place over several meetings, utilizing multiple strategies to generate signal words, including sourcing cues used in related work (e.g., Bertin et al., 2016; Chen et al., 2018), expanding this list with synonyms from online thesauruses, and ranking them by their frequency among citation sentences. This inductive process included several rounds of deliberation, manual annotation, and tests of inter-rater reliability in order to generate a robust list of candidate signal terms. The terms are intended to have high validity, but are not considered comprehensive.

We queried the database for citances containing each of these signal terms (case insensitive), using wildcards to provide for possible variants of terms (e.g., "challenge", "challenged", and "challenges"), excluding generic negation phrases ("no", "not", "cannot", "nor" and "neither" to exclude phrases such as "no conflict"), and for some signal terms excluding citances containing words associated with disciplinary jargon or methods, such as for the signal term "disagreement", which often appears with Likert-scale descriptions (e.g., "scale", "agreement", or "kappa") for survey-heavy fields. The modifications for the signal terms were derived after several rounds of review and validation. In total, citances returned by signal phrase queries comprise 3.10% (n = 145,351,937) of the database, though their relative occurrence varied dramatically, with the most coming from the "*differ**" signal term, and the least from "*disprove**" (see Table 1).

*Table 1. Specific terms comprising each of the thirteen signal term sets and specific exceptions. The "*" symbol (wildcard) captures possible variants.*

| Signal term | Variants | Exclusions | Results |
|---|---|---|---|
| **challenge*** | | | 405,613 |
| **conflict*** | | | 212,246 |
| **contradict*** | | | 115,375 |
| **contrary** | | | 171,711 |
| **contrast*** | | | 1,257,866 |
| **controvers*** | | | 154,608 |
| **debat*** | | "parliament* debat*", "congress* debat*", "senate* debat*", "polic* debat*", "politic* debat*", "public* debat*", "societ* debat*" | 150,617 |
| **differ*** | | "different*" | 2,003,677 |
| **disagree*** | "not agree*", "no agreement" | "range", "scale", "kappa", "likert", "agree*" and/or "disagree" within a ten-word range of each other. | 52,615 |
| **disprov*** | | "prove*" and "disprove*" within a ten-word range | 2,938 |
| **no consensus** | "lack of consensus" | "consensus sequence", "consensus site" | 16,632 |
| **questionable** | | | 24,244 |
| **refut*** | | "refutab*" | 10,322 |
| **total** | | | 4,578,464 |

In order to more precisely capture valid instances of disagreement and to understand their function within the literature, we also queried for citances containing both the signal terms along with at least one of four sets of *filter* terms, with no more than four words separating signal and filter. As with signal terms, filter terms were derived from iterative manual efforts of the authors to identify terms most associated with valid instances of disagreement. Four distinct sets of terms were identified, corresponding to explicit mentions of terms relating to past studies, ideas, methods, and results (see Table 2). As with signal phrases alone, the relative incidence of signal and filter phrase combinations varies widely (Table SI 1). Queries were constructed for each combination of signal term (13 total) and filter term (4 total), producing 52 combined queries, alongside 13 queries consisting only of standalone signal terms unrestricted by filter terms, for a total of 65 queries.

*Table 2. Specific terms comprising each of the four filter term sets*

| | |
|---|---|
| **studies** | studies; study; previous work; earlier work; literature; analysis; analyses; report; reports |
| **ideas** | idea*; theory; theories; assumption*; hypothesis; hypotheses |
| **methods** | model*, method*, approach*; technique* |
| **results** | result*; finding*; outcome*; evidence; data; conclusion*; observation* |

## Query Validation

From each set of results returned by the 65 queries, we selected 50 sentences for validation using simple random sampling without replacement (only 40 citances existed for "no consensus" +"ideas"), resulting in over 3,000 queried sentences. For each query, two coders were randomly selected from among the seven authors on this paper to manually annotate each citance as a valid or invalid instance of disagreement. The label was chosen based only on the text in the citation sentence, without knowledge on the citing paper's title, authors, field of study, or the surrounding text.

Consider the following four example sentences listed below (where (…) indicates the position of cited references and […] indicates additional text not quoted here). The first is invalid because the signal term, "conflict", refers to an object of study, and not a scientific dispute; the second sentence is also invalid because the term "conflicting" refers to results within a single study, not between studies; the third sentence is invalid because "challenge" appears while quoting the cited study; the fourth and fifth sentence are both examples of sentences that would be marked as valid. Similar patterns can be observed for other signal terms, such as *challenge** (Table SI 2).

1. **Invalid:** "To facilitate **conflict** management and analysis in Mcr (…), the Graph Model for Conflict Resolution (GMCR) (…) was used."
2. **Invalid:** "The 4-year extension study provided ambiguous […] and **conflicting** post hoc […] results."
3. **Invalid:** "Past studies (...) review the theoretical literature and concludes that future empirical research should '**challenge** the assumptions and analysis of the theory'."
4. **Valid:** "These observations are rather in **contradiction** with Smith et al.'s […]."
5. **Valid:** "Although there is substantial evidence supporting this idea, there are also recent **conflicting** reports (…)."

We assessed the labels for each signal/filter term combination with two measures: percent agreement (% agree) and percent valid (% valid). Percent agreement is the proportion of annotated citances in which both coders agreed on the same label of valid or invalid; this measure provides a simple measure of coder's consensus. Here, percent agreement is justified over more complicated measures such as Cohen's kappa due to the small sample of data per signal/filter term combination, and given that there are only two categories and coders.

Most signal/filter term combinations had high agreement (Figure 1a). The overall percentage agreement between coders was high, at 85.5 percent (and Cohen's kappa of 0.66). Given the difficulty of interpreting academic texts, this high percentage agreement demonstrates the robustness of our operationalization of disagreement. The signal term with the highest average agreement was *no consensus* (95.8 percent). There were only a few combinations with very low percentage agreement, mostly regarding the signal term *questionable,* which had an average lowest average percent agreement (64 percent); the nature of sentences returned from the *questionable* keyword tended to constitute marginal cases of disagreement. There was virtually no variance between the average percent agreement aggregated across filter terms. However, certain combinations of signal and filter terms were notable in resulting in higher or lower performance. For example, the difference between the highest agreement, *differ* _standalone_ (*100 percent*),* and *differ* +methods* (74 percent) is 26 points—the addition of filter terms can dramatically impact the kinds of citances returned by the query.

We calculate the percent valid as the percentage of citances annotated as valid by both coders; this provides an intuitive measure of the validity and reliability of a query. Signal/filter term combinations that best capture disagreement should have both high percent agreement and high percent validity. Not all signal/filter term combinations were found to be sufficiently valid (Figure 1b). Overall, 61.6 percent of all citances were coded as valid, with large variance between the most valid (100 percent), and the least valid (0 percent) combinations. The signal term with the highest average validity regardless of filter term was *no consensus* (94.9 percent), followed by *controvers** (88.8 percent) and *debat** (82.4 percent). Unlike with percent agreement, average validity differs drastically between filter terms, with all having higher average validity than *_standalone_*. The combinations with highest validity are *no consensus + studies* (98 percent), *no consensus +methods* (98 percent), and *no consensus _standalone_* (94 percent) For specific signal terms, the presence of a filter term can have a drastic impact of coded validity; for example, the validity of *contrast** +ideas (80 percent) is four times greater than of *contrast* _standalone_* and *contrast +methods (20 percent).*

The queries that best capture instances of disagreement are those with the highest validity. We choose a validity threshold of 80 percent and exclude queries with lower validity from subsequent analysis. We also consider several adjustments to the threshold to assess the robustness of our empirical findings. 23 queries sit above this the 80 percent threshold (Figure 1c), including all five *no consensus* and *controvers** queries, four *debat** queries, two *disagree** and *contradict** queries, and one query each for *contrary**, *contrast**, *conflict**, *disprove*,* and *questionable*. Because we prioritized precision, these 23 queries comprise only a fraction of total citances: 455,625, representing 0.31% of all citances in our dataset. We note that citances returned by queries are not exclusive; for example, a citance containing both *controvers** and *no consensus** would count towards both signal phrases. Similarly, a citance returned with the query *controvers* +methods* would also be returned by the *controvers**. Naturally, more general queries, such as *differ** and *contrast** returned a much greater number of citances. Among queries above the 80 percent threshold, the *controvers** and *debat** produce the highest number of citances (154,608 and 150,617 respectively, Figure 1d). The 455,625 citances returned by our queries as well as relevant publication and query details are available in Zenodo (Lamers & Van Eck, 2021).

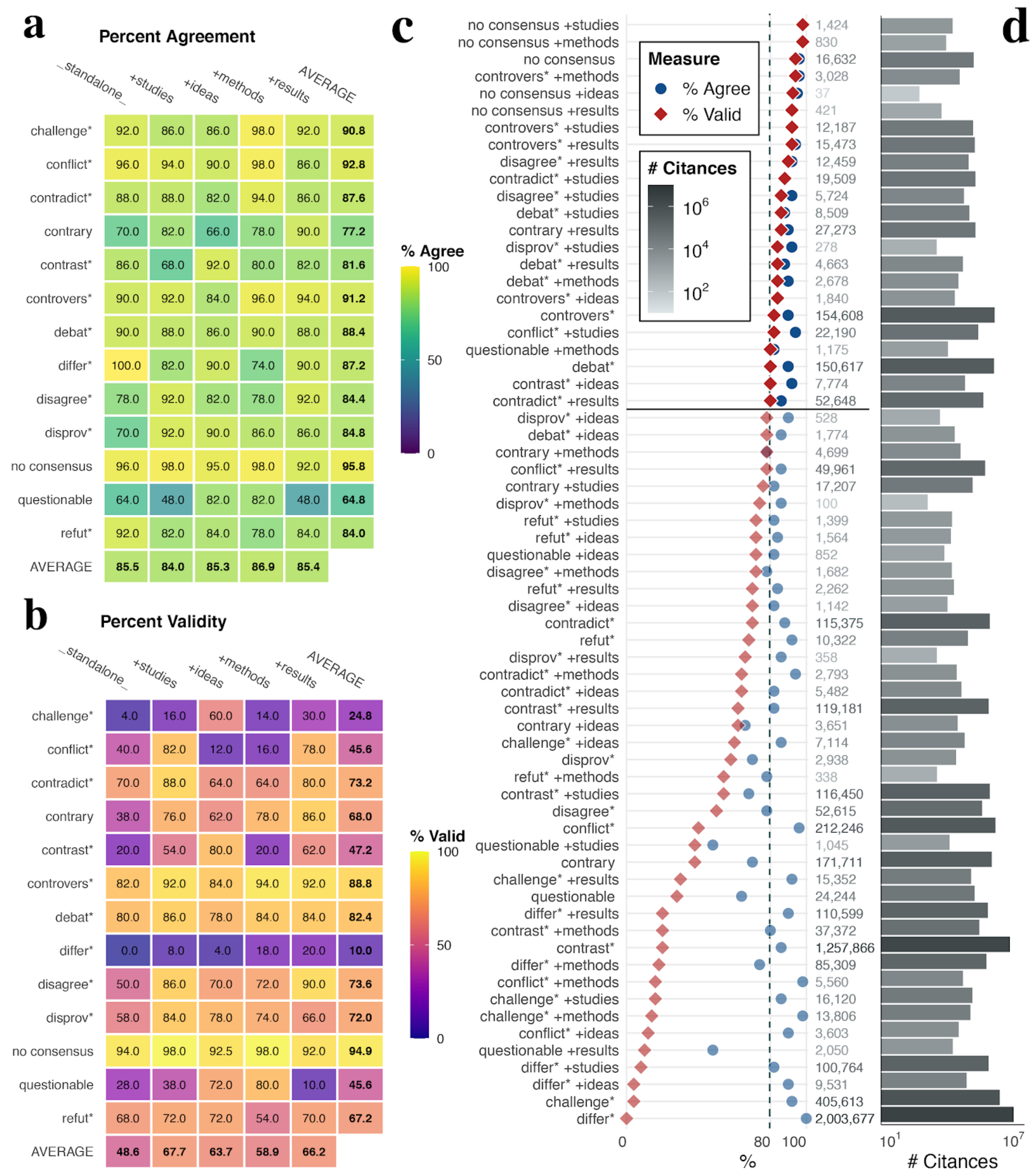

*Figure 1, Agreement and validity of signal and filter term combination. Measures calculated from 50 randomly-sampled citances for each combination annotated as valid or invalid instances of disagreement by two independent coders. a. Percentage agreement, or the proportion of citances for which coders independently agreed on the label. b. Percentage validity, or the proportion of citances which both coders labeled as valid. Averages are shown by row and column. c. Percentage agreement and validity of each signal/filter term combination, ordered from highest percent validity (top) to lowest percent validity (bottom). Numbers on the right are the total number of citances returned by querying using the signal/filter term combination, and are colored according to their log-transformed value. d. Log-transformed count of citances returned by each query combination, colored by the (log-transformed)*

*number of citances. Citance counts are non-exclusive, meaning that citances of the form debat* +studies will also be counted towards debat* _standalone_.*

As a confirmation of overall validity, we measure the rate of disagreement by instances of self-citation and non-self-citation. We expect that authors will be less likely to cite their own work within the context of disagreement. Indeed, we find that the rate of disagreement for non-self-citations is 2.4 times greater than for self-citations (Figure SI 3), demonstrating that our indicator of disagreement affirms expectations.

# Results

Instances of disagreement, operationalized using the 23 validated queries, accounted for approximately 0.31 percent of all citation sentences (citances) extracted from indexed papers published between 2000 and 2015 (Figure 2a). Disagreement was highest in the Social Sciences and Humanities (Soc & Hum, 0.61 percent), followed by Biomedical and Health Sciences (Bio & Health, 0.41 percent), Life and Earth Sciences (Life & Earth, 0.29 percent), Physical Sciences and Engineering (Phys & Engr, 0.15 percent), and Mathematics and Computer Science (Math & Comp, 0.06 percent).

Our measure shows that disagreement has been relatively constant over time (Figure 2b), decreasing at an average rate of about 0.0005 percentage points per year. This is driven by falling disagreement in Phys & Engr (-0.0045 points per year), Soc & Hum (-0.0033 points per year), and Math & Comp (-0.0019 points per year). Phys & Engr stands out not only for its stable decrease each year, but also for its relative size; given a starting rate of one disagreement signal per 529 citances in 2000, by 2015 the rate of disagreement in Physics fell to one disagreement per 809 citances, a 35 percent decrease, compared to a 24 percent decrease for Math & Comp and only a 5 percent decrease in Soc & Hum. In contrast, disagreement has tended to increase somewhat in Bio & Health (+0.0017 points per year) and Life & Earth (+0.0018 points per year). These trends are likely not the result of uses of individual queries; for example, *disagree** queries are over-represented in Phys & Engr (Figure SI 2), yet the incidence of these terms is falling or remaining stable (Figure SI 1). Similarly, *debat** was over-represented in Soc & Hum and has increased in usage despite slight falling disagreement in the field. That these changes are not confined to any single query suggests that field-level differences represent changes in the level of disagreement within a field rather than linguistic or methodological artifacts. These findings are also consistent at a lower, 70% validity threshold for disagreement queries, which includes 13 new queries that bring the total number of disagreement citances to over 650,000 (see Supporting Information).

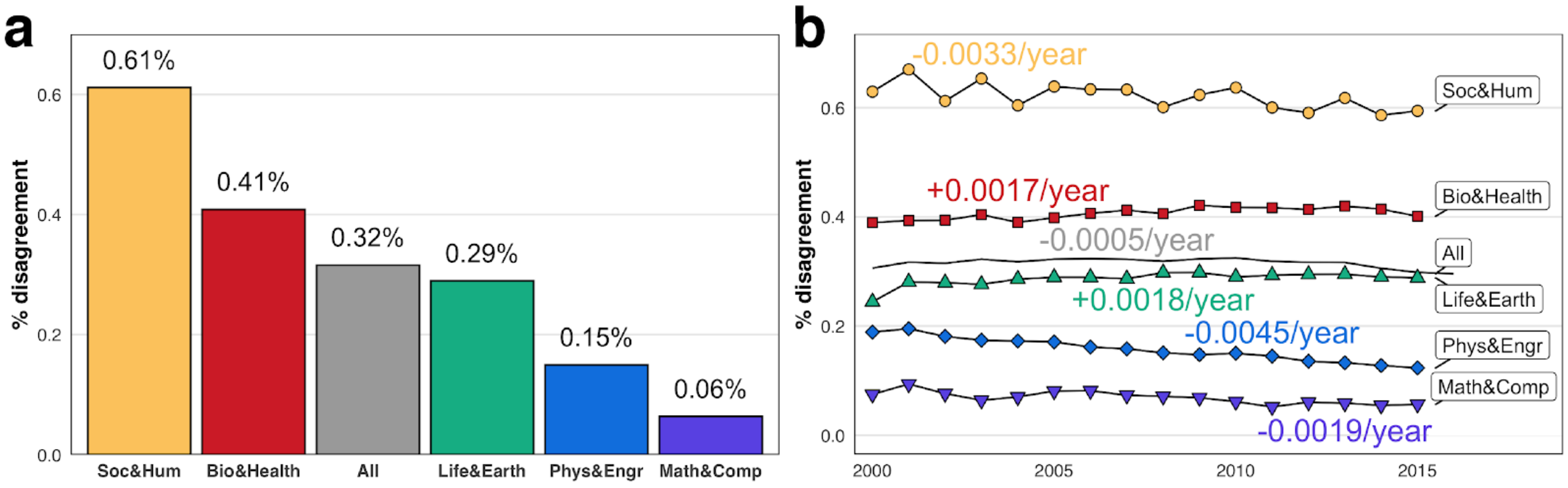

***Figure 2**. Disagreement reflects a hierarchy of fields. (a) Percent of all citances in each field that contain signals of disagreement, meaning they were returned by one of the 23 queries with validity of 80 percent or higher. Fields marked by lower consensus, such as in Soc & Hum, had a greater proportion of disagreement. (b) Percent of disagreement by field and over time, showing little change overall, but some changes by field. Text indicates the average percentage-point change per-year by field.*

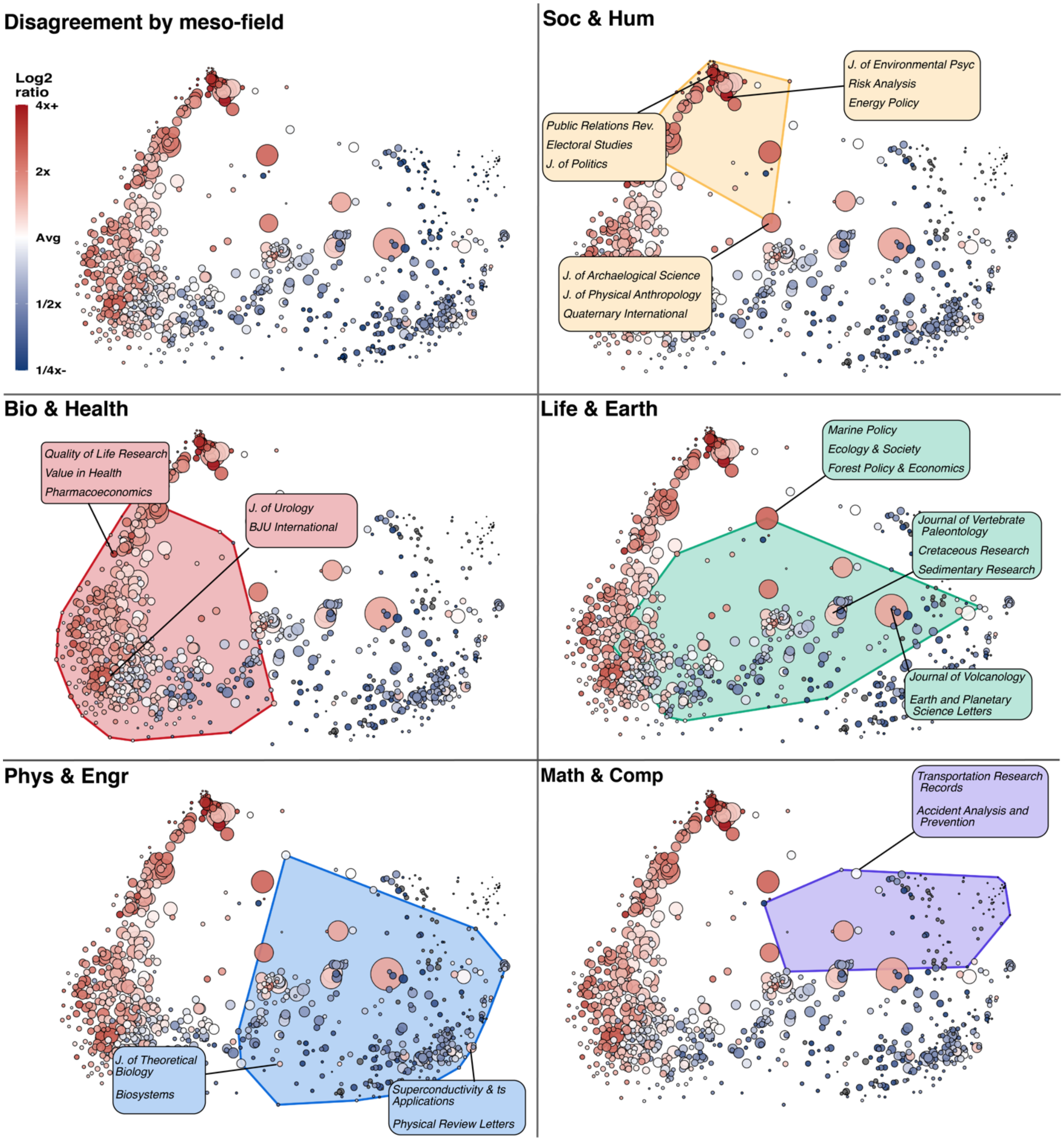

***Figure 3. Heterogeneity in disagreement across meso-fields**. Fine-grained view across 817 meso-level fields, each a cluster of publications grouped and positioned based on their citation links derived from the Web of Science database (see Materials & Methods), 2000-2015. The area of each point is proportional to the number of disagreement citances in that field. Overlapping points are an artifact of their position and size, and bear no additional meaning. Color maps to the log ratio of the share of disagreement citances given the mean share across all fields, truncated at 4x greater and 4x lower than the mean. Soc & Hum tends to have a greater proportion of disagreement citances, and Math & Comp the least. Other panels show the same data, but highlight the meso-fields in each high-level field. Meso-fields of interest are highlighted, and labels show a selection of journals in which papers in each field are published. Journals listed in labels are representative of each meso-field in the Web of Science, and is not limited to those represented in the Elsevier ScienceDirect data. An interactive version of this visualization is available online at https://tinyurl.com/disagreement-meso-fields.*

The more fine-grained meso-fields reveal heterogeneity within the larger fields (Figure 3). Overall, meso-field disagreement followed the same pattern as Figure 2, with higher scores in Soc & Hum and lower in Math & Comp. However, some meso-fields stand out. For example, some of the highest rates of disagreement found in the Bio & Health meso-field was in more *social* journals such as *Quality of Life Research, Value in Health,* and *Pharmacoeconomics*. Similarly, in Math & Comp, the meso-field with the most disagreement contained journals relating to transportation science, a technical field which draws on management studies and other social science literature. This pattern held in Life & Earth, in which a meso-field with a relatively high share of disagreement contained papers in journals such as *Marine Policy*, *Ecology & Society*, and *Forest Policy & Economics.* The high disagreement in these meso-fields lends support to the hypothesis that regardless of the high-level field, more socially-oriented topics generate a higher level of disagreement. Also notable is that, in Life & Earth, several large fields with relatively high disagreement study the distant geological past or other inaccessible objects of studies, comprised of papers in journals such as the *Journal of Vertebrate Paleontology*, *Cretaceous Research*, and *Sedimentary Research*. A similar observation can be made in Phys & Engr, where astronomy-related fields featuring journals such as *Planetary and Space Science* and *Theoretical Biology* exhibit above-average rates of disagreement, along with fields pertaining to research into superconductivity. Field-level results must be interpreted cautiously, however, as our signal terms may mis-classify citances based on disciplinary keywords and jargon (see Supporting Information).

In addition to these quantitative results, we perform a qualitative investigation of the individual papers that issued the most disagreement citations, and which were cited most often in the context of disagreement. First, we examine the citing paper perspective, that is those papers that issued the most citances (Table SI 6). These top papers demonstrate how methodological artifacts can contribute to these more extreme examples. For example, one of these papers considers the pedagogical and evaluative potential of debates in the classroom (Doody & Condon, 2012, Table SI 6); the *"debat*"* signal term incorrectly classifies several citations included in this paper as evidence of scientific disagreement. However, other papers offer interesting instances of disagreement, and exemplify lessons that should be considered when quantifying disagreement. For instance, one such paper concerns meteorite impact structures (French & Koeberl, 2010, Table SI 6) and includes discussion on the controversies in the field. Another is a review article arguing for multi-target agents for treating depressive states (Millan, 2006, Table SI 6), and catalogs the controversies around the topic. Yet another is a book on *Neurotoxicology and Teratology*, misclassified as a research article, and illustrates how the length of an article can contribute to its likelihood of issuing a disagreement citation (Kalter, 2003, Table SI 6).

Considering the cited paper perspective—those papers that received the most paper-level disagreement citations or were referenced the most in the context of community disagreement—reveals clear instances of disagreement in the literature. Many of the studies receiving the most disagreement citances (Table SI 7) relate to a single longstanding scientific controversy in the Earth sciences concerning the formation of the North Chinese Cranton, a tectonic structure spanning Northern China, Inner Mongolia, the Yellow Sea, and North Korea. This list of most-disagreed-with papers also includes a literature review that is cited as an exemplar of controversy, here regarding the existence of "lipid rafts" in cells (Munro, 2003, Table SI 7), and a paper on fMRI research that is heralded as a methodological improvement in the field, and is often cited to draw a contrast with other methods (Murphy et al., 2009, Table SI 7). A more thorough discussion of papers that issue and receive the most disagreement can be found in the Supporting Information.

We also investigated the extent to which other factors relate to a paper's being cited in the context of disagreement. Younger papers (relative to the citing paper) are more likely to receive a disagreement citance than older ones (Figure SI 4), which may indicate that the role of cited literature varies based on its age (He & Chen, 2018). Author demographics do not appear to play a strong role; here, we observe little difference in disagreement based on the gender of the citing author (Figure SI 6). We also explore whether disagreement relates to citation impact; whereas previous analysis revealed a positive relationship between conflict and citation (Radicchi, 2012), our preliminary results do not find evidence of increased citation, at least in the years immediately following the disagreement (See Supporting Information). Papers that themselves contain disagreement citances, however, tend to receive more citations over their lifespan.

# Discussion

When it comes to defining scientific disagreement, scholars disagree. Rather than staking out a specific definition, we adopt a broad operationalization of disagreement that incorporates elements of Kuhn's accumulation of anomalies (Kuhn, 1996), Latour's controversies (Latour, 1988), and more recent notions of uncertainty (Chen et al., 2018) and negative citations (Catalini et al., 2015). By bridging these past theories, we quantify the rate of disagreement across science. Roughly 0.31 percent of all citances in our dataset are instances of disagreement, a share that has remained relatively stable over time. However, this number is much smaller than in past studies—such as the 2.4 percent for so-called "negative" references (Catalini et al., 2015), and the estimated 0.8 percent for "disputing" citations (Nicholson et al., 2021). This is explained by our operationalization of disagreement, which although conceptually broader than negative or disputing citations, is narrowed to only 23 queries to prioritize precision. Moreover, studies differ in corpus used, most often covering only one journal or field, compared to our large multidisciplinary corpus. The strength of our analysis is not the absolute incidence of disagreement, but its relative differences across disciplinary and social contexts.

Disagreement across fields can be interpreted using several theoretical frameworks. Differences in disagreement might stem from the *epistemic* characteristics of fields and their topics of study. For example, Auguste Comte's *hierarchy of sciences* model (Comte, 1856) proposed that fields are organized based on the inherent complexity of their subject matter. We reinforce this model, finding that disagreement is highest in fields at the top of the hierarchy, such as the social sciences and humanities, and lowest in fields at the bottom of the hierarchy, such as physics and mathematics. While the hierarchy of sciences model is well-grounded theoretically (Cole, 1983) and bibliometrically (Fanelli, 2010; Fanelli & Glänzel, 2013), other frameworks may be equally useful in understanding disagreement across fields. For example, the *structural* characteristics of fields may explain their differences in disagreement. One such characteristic is how reliant the field is on Kuhnian paradigms (Kuhn, 1996); so-called "hard" sciences, such as physics, may have strong theoretical paradigms and greater consensus (less disagreement) than "soft" sciences such as those in the social sciences and humanities (Biglan, 1973). Changes in these structural characteristics may also contribute to the temporal evolution of disagreement. For instance, the decrease of disagreement in physics and engineering may be due to a transition into a period of "normal" science (Kuhn, 1996), as it has been previously argued for certain sub-fields (Smolin, 2007). Increase in collaboration (Wuchty et al., 2007) may also affect the trends, as consensus has to be reached among a larger body of individuals during the research process. Social sciences and humanities have other characteristics that might be associated with more common or more intense conflicts, including low centralization of resources and control over research agendas, high diversity in their audiences and stakeholders, and limited standardization of methods and theories (Whitley, 2000). A field's *cultural* characteristics also play a role in its norms of disagreement. Fields have different norms when it comes to consensus formation and the settling of disputes (Cetina, 1999), and some fields even value disagreement as an important element of scholarship. For instance, a cultural norm of "agonism", or ritualized adversarialism, is common in many humanities fields, wherein one's arguments are framed in direct opposition to past arguments (Tannen, 2002). Fields also have distinct cultures of evaluation, which shapes how they judge each other's work and impacts whether they are likely to reach consensus (Lamont, 2009). Of course, epistemic, structural, and cultural characteristics of fields are all inter-related—cultural practices emerge in part from structural characteristics of a field, such as access to expensive instruments, which in turn are related to the epistemic aspects of the object of study.

Our data does not allow us to disentangle these relationships or argue which is most appropriate, but each offers a useful lens for understanding why disagreement might differ between fields.

Expanding our analysis into a more fine-grained classification of fields reveals greater detail into where disagreement happens in science. We observed that socially-oriented meso-level fields tended to have a higher rate of disagreement, no matter their main field. For example, meso-fields concerning healthcare policy had higher rates of disagreement than others in Biomedical & Health Sciences, whereas the meso-field concerning transportation science had a higher rate of disagreement than all others in Math & Computer science. Though these fields draw on the expertise of traditionally hard-science fields, they do so in order to study social processes and address social questions. In Life & Earth Sciences, disagreement was especially high in meso-fields that study the earth's geological and paleontological history. In these fields, much like in the social sciences, researchers cannot easily design experiments, and so progress instead comes from debate over competing theories using limited evidence and reconstructed historical records. This is exemplified by paleontology, in which a 2017 paper sparked controversy and forced a re-interpretation of the fossil record and a 130-year-old theory of dinosaur evolution (Baron et al., 2017; Langer et al., 2017). Similarly, our approach identified a major controversy in the earth sciences—the formation of the North Chinese Cranton—again illustrating how reliance on historical records might exacerbate disagreement. These cases illustrate the heterogeneity of disagreement in science, and illustrate that existing theoretical frameworks, such as the hierarchy of science, can oversimplify the diversity of cultural norms and epistemic characteristics that manifest at more fine-grained levels of analysis.

Our approach comes with limitations. First, our method captures only a fraction of textual disagreements in science. This is partly due to our prioritization of precision over recall, having removed cue-phrases with low validity. Our lists of signal and filter terms are also non-exhaustive, and so their extension in future research would identify more instance of disagreement. Given our focus on citances, we are not able to identify traces of disagreements that occur without explicit reference to past literature, or those that can only be classified as disagreement with surrounding sentences as context. Some disagreements may also be too subtle, or rely on technical jargon, such that they cannot be identified with our general signal terms. Moreover, our measure does not capture non-explicit disagreements, or scientific disagreements occurring outside of citances, such as in conferences, books, social media, or in interpersonal interactions. For these reasons, our measure of disagreement may over- or under-represent disagreement in particular fields, and should be considered when evaluating results. Second, in spite of its overall precision, our approach returns many false positives in particular disciplinary contexts. For example, the signal term *conflict** matches to topics of study and theories in the fields of sociology and international relations (e.g., "ethnic conflicts", "Conflict theory"). In other instances, a signal term can even match an author's name (as in the surname "Debat"). We also find that these artifacts are over-represented among the papers that issued the most disagreement citances, and those that were most often cited in the context of disagreement (see Supporting Information). However, given our extensive validation, these artifacts remain a small minority of all disagreement sentences identified, though they should be considered when interpreting disagreement in small sub-fields. Finally, our inclusive definition of disagreement homogenizes disagreement into a single category, whereas there are many kinds of disagreement in science. For example, the ability to differentiate between *paper-level* and *community*-level disagreement could lend insight into how conflict and controversy manifest in different fields. This definition could also be developed to differentiate further between types of disagreement: for example, past citation classification schemes have differentiated between "juxtaposition" and "negational" citations (Moravcsik & Murugesan, 1975), or between "weakness" and "contrast" citations (Teufel et al., 2006).

Despite these limitations, our framework and study have several advantages. First, in contrast to keyword-based analyses, our approach provides a nuanced view of disagreement in science, revealing the differences in disagreement not only between signal terms, but also based on filter terms. This drives the second advantage of our approach—that its inherent *transparency* allows us to easily identify confounding artifacts such as when a signal term is an object of study (i.e., “international *conflicts*”, “public *debate”)*, when it relates to disciplinary jargon (i.e., “*disproving* theorems” in Mathematics, or *“strongly disagree”* in survey studies that use Likert scales) or when the keyword is part of a proper name (i.e., “work by *Debatin* et al.,”). These issues are a concern for any automated analysis of scientific texts across disciplines—the usage and meaning of words varies across fields. In contrast to black-box style machine learning approaches, ours is transparent and can easily be validated, interpreted, replicated, and extended. Finally, by being open and transparent, our approach is easily adjustable to different contexts. Our initial identification of keywords was the result of an iterative process of exploration and validation, which eventually resulted in a non-exhaustive set of signal terms, filter terms, exclusions, and then a final set of validated queries. Any step of this process can be tuned, extended, and improved to facilitate further studies of scientific disagreement—new signals or filters can be introduced, queries can be modified to be even more precise, and the threshold of validity changed; here, for example, we assessed our results by setting a more inclusive threshold for which queries constitute disagreement, and find the results remain robust (see Supporting Information). To assist in further efforts to validate and extend our work, we have made annotated sentences and code that can reproduce this analysis publicly available at *github.com/murrayds/sci-text-disagreement.*

Whereas black-box machine learning approach have many strengths (e.g. Rife et al., 2021), ours is transparent and intuitive. Its transparency allows to easily identify terms that have field-specific meanings, which may be obfuscated in black-box approaches. Our approach is also reproducible and can be refined and extended with additional signal and filter terms. The portability of our queries also mean that they can readily be applied to other full-text data. The general method of generating and manually validating signal and filter terms can also be applied to other scientific phenomena, such as detecting uncertainty (Chen et al., 2018), negativity (Catalini et al., 2015), discovery (Small et al., 2017), or an expanded framework of disagreement (Moravcsik & Murugesan, 1975).

Future research could refine and extend these existing queries and link them to different conceptual perspectives on disagreement in science. Such work could build on our analyses of the factors that may affect disagreement, including gender, paper age, and citation impact. Disagreement is an essential aspect of knowledge production, and understanding its social, cultural, and epistemic characteristics will reveal fundamental insights into science in the making.

# Acknowledgements

We thank Yong-Yeol Ahn, Staša Milojević, Alessandro Flammini, Filippo Menczer, Dashun Wang, Lili Miao, and participants of the *A scientometric analysis of disagreement in science* seminar held at CWTS at Leiden University for their helpful comments. We are grateful to Elsevier for making the full-text data available to us.

# Author Contributions

Wout S. Lamers: Conceptualization; Formal analysis; Investigation; Methodology; Software; Visualization; Writing – original draft; Writing – review & editing
Kevin Boyack: Conceptualization; Investigation; Methodology; Writing – original draft; Writing – review & editing
Vincent Larivière: Conceptualization; Investigation; Methodology; Writing – review & editing
Cassidy R. Sugimoto: Conceptualization; Investigation; Methodology; Writing – review & editing
Nees Jan van Eck: Conceptualization; Data curation; Formal analysis; Investigation; Methodology; Software; Visualization; Writing – review & editing
Ludo Waltman: Investigation; Methodology; Writing – review & editing
Dakota Murray: Conceptualization; Formal analysis; Investigation; Methodology; Project administration; Software; Visualization; Writing – original draft; Writing – review & editing

# Funding Information

Dakota Murray is supported by the Air Force Office of Scientific Research under award number FA9550-19-1-039. Vincent Larivière acknowledges funding from the Canada Research Chairs program.

# Data Availability Statement

The Elsevier full-text data and the Web of Science bibliographic data used in this study were obtained from proprietary data sources. We are not allowed to share the raw data on which our study is based. We do have permission from Elsevier to share a data set containing the 455,625 citing sentences identified using our disagreement queries. This data set is available in Zenodo (Lamers & Van Eck, 2021).

# References

Alén, E., Domínguez, T., & de Carlos, P. (2015). University studentsʹ perceptions of the use of academic debates as a teaching methodology. *Journal of Hospitality, Leisure, Sport & Tourism Education*, *16*, 15–21. https://doi.org/10.1016/j.jhlste.2014.11.001

Balietti, S., Mäs, M., & Helbing, D. (2015). On Disciplinary Fragmentation and Scientific Progress. *PLOS ONE*, *10*(3), e0118747. https://doi.org/10.1371/journal.pone.0118747

Baron, M. G., Norman, D. B., & Barrett, P. M. (2017). A new hypothesis of dinosaur relationships and early dinosaur evolution. *Nature*, *543*(7646), 501–506. https://doi.org/10.1038/nature21700

Bertin, M., Atanassova, I., Sugimoto, C. R., & Lariviere, V. (2016). The linguistic patterns and rhetorical structure of citation context: An approach using n-grams. *Scientometrics*, *109*(3), 1417–1434. https://doi.org/10.1007/s11192-016-2134-8

Biglan, A. (1973). The characteristics of subject matter in different academic areas. *Journal of Applied Psychology*, *57*(3), 195–203. https://doi.org/10.1037/h0034701

Bornmann, L., Wray, K. B., & Haunschild, R. (2020). Citation concept analysis (CCA): A new form of citation analysis revealing the usefulness of concepts for other researchers illustrated by exemplary case studies including classic books by Thomas S. Kuhn and Karl R. Popper. *Scientometrics*, *122*(2), 1051–1074. https://doi.org/10.1007/s11192-019-03326-2

Boyack, K. W., van Eck, N. J., Colavizza, G., & Waltman, L. (2018). Characterizing in-text citations in scientific articles: A large-scale analysis. *Journal of Informetrics*, *12*(1), 59–73. https://doi.org/10.1016/j.joi.2017.11.005

Bruggeman, J., Traag, V. A., & Uitermark, J. (2012). Detecting Communities through Network Data. *American Sociological Review*, *77*(6), 1050–1063. https://doi.org/10.1177/0003122412463574

Bruschke, J., & Divine, L. (2017). Debunking Nixon's radio victory in the 1960 election: Re-analyzing the historical record and considering currently unexamined polling data. *The Social Science Journal*, *54*(1), 67–75. https://doi.org/10.1016/j.soscij.2016.09.007

Castelvecchi, D. (2020). Mystery over Universe's expansion deepens with fresh data. *Nature*, *583*(7817), 500–501. https://doi.org/10.1038/d41586-020-02126-6

Catalini, C., Lacetera, N., & Oettl, A. (2015). The incidence and role of negative citations in science. *Proceedings of the National Academy of Sciences*, *112*(45), 13823–13826. https://doi.org/10.1073/pnas.1502280112

Cetina, K. K. (1999). *Epistemic Cultures: How the Sciences Make Knowledge*. Harvard University Press.

Chen, C., Song, M., & Heo, G. E. (2018). A Scalable and Adaptive Method for Finding Semantically Equivalent Cue Words of Uncertainty. *Journal of Informetrics*, *12*(1), 158–180. https://doi.org/10.1016/j.joi.2017.12.004

Cole, S. (1983). The Hierarchy of the Sciences? *American Journal of Sociology*, *89*(1), 111–139. https://doi.org/10.1086/227835

Cole, S., Simon, G., & Cole, J. R. (1988). Do Journal Rejection Rates Index Consensus? *American Sociological Review*, *53*(1), 152–156. https://doi.org/10.2307/2095740

Collins, H. (2017). *Gravity's Kiss: The Detection of Gravitational Waves* (1st edition). The MIT Press.

Colston, N. M., & Vadjunec, J. M. (2015). A critical political ecology of consensus: On “Teaching Both Sides” of climate change controversies. *Geoforum*, *65*, 255–265. https://doi.org/10.1016/j.geoforum.2015.08.006

Comte, A. (1856). *The Positive Philosophy of Auguste Comte*. Calvin Blanchard.

Debat, P., Nikiéma, S., Mercier, A., Lompo, M., Béziat, D., Bourges, F., Roddaz, M., Salvi, S., Tollon, F., & Wenmenga, U. (2003). A new metamorphic constraint for the Eburnean orogeny from Paleoproterozoic formations of the Man shield (Aribinda and Tampelga countries, Burkina Faso). *Precambrian Research*, *123*(1), 47–65. https://doi.org/10.1016/S0301-9268(03)00046-9

Dieckmann, N. F., & Johnson, B. B. (2019). Why do scientists disagree? Explaining and improving measures of the perceived causes of scientific disputes. *PLOS ONE*, *14*(2), e0211269. https://doi.org/10.1371/journal.pone.0211269

Doody, O., & Condon, M. (2012). Increasing student involvement and learning through using debate as an assessment. *Nurse Education in Practice*, *12*(4), 232–237. https://doi.org/10.1016/j.nepr.2012.03.002

Ersoy, A. F. (2010). Social studies teacher candidates’ views on the controversial issues incorporated into their courses in Turkey. *Teaching and Teacher Education*, *26*(2), 323–334.

Evans, E., Gomez, C., & McFarland, D. (2016). Measuring Paradigmaticness of Disciplines Using Text. *Sociological Science*, *3*, 757–778. https://doi.org/10.15195/v3.a32

Fanelli, D. (2010). “Positive” Results Increase Down the Hierarchy of the Sciences. *PLOS ONE*, *5*(4), e10068. https://doi.org/10.1371/journal.pone.0010068

Fanelli, D., & Glänzel, W. (2013). Bibliometric Evidence for a Hierarchy of the Sciences. *PLOS ONE*, *8*(6), e66938. https://doi.org/10.1371/journal.pone.0066938

French, B. M., & Koeberl, C. (2010). The convincing identification of terrestrial meteorite impact structures: What works, what doesn't, and why. *Earth-Science Reviews*, *98*(1), 123–170. https://doi.org/10.1016/j.earscirev.2009.10.009

Hargens, L. L. (1988). Scholarly Consensus and Journal Rejection Rates. *American Sociological Review*, *53*(1), 139–151.

He, J., & Chen, C. (2018). Temporal Representations of Citations for Understanding the Changing Roles of Scientific Publications. *Frontiers in Research Metrics and Analytics*, *3*, 27. https://doi.org/10.3389/frma.2018.00027

Hyland, K. (1998). *Hedging in Scientific Research Articles* (Vol. 54). John Benjamins Publishing Company. https://doi.org/10.1075/pbns.54

Kalter, H. (2003). Teratology in the 20th century: Environmental causes of congenital malformations in humans and how they were established. *Neurotoxicology and Teratology*, *25*(2), 131–282. https://doi.org/10.1016/S0892-0362(03)00010-2

Kitcher, P. (1995). *Advancement of Science: Science Without Legend, Objectivity Without Illusions*. Oxford University Press.

Kuhn, T. S. (1996). *The Structure of Scientific Revolutions* (3rd edition). University of Chicago Press.

Kusky, T. M. (2011). Geophysical and geological tests of tectonic models of the North China Craton. *Gondwana Research*, *20*(1), 26–35. https://doi.org/10.1016/j.gr.2011.01.004

Kusky, T. M., & Li, J. (2003). Paleoproterozoic tectonic evolution of the North China Craton. *Journal of Asian Earth Sciences*, *22*(4), 383–397. https://doi.org/10.1016/S1367-9120(03)00071-3

Lamers, W. S., & Van Eck, N. J. (2021). *Measuring Disagreement in Science* [Data set]. Zenodo. https://doi.org/10.5281/zenodo.5148058

Lamont, M. (2009). *How Professors Think: Inside the Curious World of Academic Judgment*. Harvard University Press.

Langer, M. C., Ezcurra, M. D., Rauhut, O. W. M., Benton, M. J., Knoll, F., McPhee, B. W., Novas, F. E., Pol, D., & Brusatte, S. L. (2017). Untangling the dinosaur family tree. *Nature*, *551*(7678), E1–E3. https://doi.org/10.1038/nature24011

Larivière, V., Ni, C., Gingras, Y., Cronin, B., & Sugimoto, C. R. (2013). Bibliometrics: Global gender disparities in science. *Nature News*, *504*(7479), 211. https://doi.org/10.1038/504211a

Latour, B. (1988). *Science in Action: How to Follow Scientists and Engineers Through Society* (Reprint edition). Harvard University Press.

Li, Z. X., Bogdanova, S. V., Collins, A. S., Davidson, A., De Waele, B., Ernst, R. E., Fitzsimons, I. C. W., Fuck, R. A., Gladkochub, D. P., Jacobs, J., Karlstrom, K. E., Lu, S., Natapov, L. M., Pease, V., Pisarevsky, S. A., Thrane, K., & Vernikovsky, V. (2008). Assembly, configuration, and break-up history of Rodinia: A synthesis. *Precambrian Research*, *160*(1), 179–210. https://doi.org/10.1016/j.precamres.2007.04.021

Millan, M. J. (2006). Multi-target strategies for the improved treatment of depressive states: Conceptual foundations and neuronal substrates, drug discovery and therapeutic application. *Pharmacology & Therapeutics*, *110*(2), 135–370. https://doi.org/10.1016/j.pharmthera.2005.11.006

Miranda, R., & Garcia-Carpintero, E. (2018). Overcitation and overrepresentation of review papers in the most cited papers. *Journal of Informetrics*, *12*(4), 1015–1030. https://doi.org/10.1016/j.joi.2018.08.006

Moravcsik, M. J., & Murugesan, P. (1975). Some Results on the Function and Quality of Citations. *Social Studies of Science*, *5*(1), 86–92. https://doi.org/10.1177/030631277500500106

Munro, S. (2003). Lipid Rafts: Elusive or Illusive? *Cell*, *115*(4), 377–388. https://doi.org/10.1016/S0092-8674(03)00882-1

Murphy, K., Birn, R. M., Handwerker, D. A., Jones, T. B., & Bandettini, P. A. (2009). The impact of global signal regression on resting state correlations: Are anti-correlated networks introduced? *NeuroImage*, *44*(3), 893–905. https://doi.org/10.1016/j.neuroimage.2008.09.036

Nam, C. W. (2014). The effects of trust and constructive controversy on student achievement and attitude in online cooperative learning environments. *Computers in Human Behavior*, *37*, 237–248. https://doi.org/10.1016/j.chb.2014.05.007

Nicholson, J. M., Mordaunt, M., Lopez, P., Uppala, A., Rosati, D., Rodrigues, N. P., Grabitz, P., & Rife, S. C. (2021). scite: A smart citation index that displays the context of citations and classifies their intent using deep learning. *BioRxiv*, 2021.03.15.435418. https://doi.org/10.1101/2021.03.15.435418

Nicolaisen, J., & Frandsen, T. F. (2012). Consensus formation in science modeled by aggregated bibliographic coupling. *Journal of Informetrics*, *6*(2), 276–284. https://doi.org/10.1016/j.joi.2011.08.001

North China Craton. (2020). In *Wikipedia*. https://en.wikipedia.org/w/index.php?title=North_China_Craton&oldid=992585145

Oreskes, N., & Conway, E. M. (2011). *Merchants of Doubt: How a Handful of Scientists Obscured the Truth on Issues from Tobacco Smoke to Climate Change* (Reprint edition). Bloomsbury Publishing.

Popper, K. (1963). *Conjectures and Refutations: The Growth of Scientific Knowledge* (2nd edition). Routledge.

Radicchi, F. (2012). In science "there is no bad publicity": Papers criticized in comments have high scientific impact. *Scientific Reports*, *2*. https://doi.org/10.1038/srep00815

Rife, S. C., Rosati, D., & Nicholson, J. M. (2021). scite: The next generation of citations. *Learned Publishing*. https://doi.org/10.1002/leap.1379

Sarewitz, D. (2011). The voice of science: Let's agree to disagree. *Nature*, *478*(7367), 7–7. https://doi.org/10.1038/478007a

Shapin, S., & Schaffer, S. (2011). *Leviathan and the Air-Pump: Hobbes, Boyle, and the Experimental Life*. Princeton University Press.

Shwed, U., & Bearman, P. S. (2010). The Temporal Structure of Scientific Consensus Formation. *American Sociological Review*, *75*(6), 817–840. https://doi.org/10.1177/0003122410388488

Shwed, U., & Bearman, P. S. (2012). Symmetry Is Beautiful. *American Sociological Review*, *77*(6), 1064–1069. https://doi.org/10.1177/0003122412463018

Small, H. (2018). Characterizing highly cited method and non-method papers using citation contexts: The role of uncertainty. *Journal of Informetrics*, *12*(2), 461–480. https://doi.org/10.1016/j.joi.2018.03.007

Small, H., Boyack, K. W., & Klavans, R. (2019). Citations and certainty: A new interpretation of citation counts. *Scientometrics, 118*(3), 1079–1092. https://doi.org/10.1007/s11192-019-03016-z

Small, H., Tseng, H., & Patek, M. (2017). Discovering discoveries: Identifying biomedical discoveries using citation contexts. *Journal of Informetrics*, *11*(1), 46–62. https://doi.org/10.1016/j.joi.2016.11.001

Smolin, L. (2007). *The Trouble With Physics: The Rise of String Theory, The Fall of a Science, and What Comes Next by Lee Smolin*. Mariner Books.

Stepanova, O., & Bruckmeier, K. (2013). The relevance of environmental conflict research for coastal management. A review of concepts, approaches and methods with a focus on Europe. *Ocean & Coastal Management*, *75*, 20–32. https://doi.org/10.1016/j.ocecoaman.2013.01.007

Szarvas, G., Vincze, V., Farkas, R., Móra, G., & Gurevych, I. (2012). Cross-Genre and Cross-Domain Detection of Semantic Uncertainty. *Computational Linguistics*, *38*(2), 335–367. https://doi.org/10.1162/COLI_a_00098

Tannen, D. (2002). Agonism in academic discourse. *Journal of Pragmatics*, *34*(10), 1651–1669. https://doi.org/10.1016/S0378-2166(02)00079-6

Teufel, S., Siddharthan, A., & Tidhar, D. (2006). Automatic Classification of Citation Function. *Proceedings of the 2006 Conference on Empirical Methods in Natural Language Processing*, 103–110. http://dl.acm.org/citation.cfm?id=1610075.1610091

The power of disagreement. (2016). *Nature Methods*, *13*(3), 185–185. https://doi.org/10.1038/nmeth.3798

Valenzuela, M., Ha, V., & Etzioni, O. (2015). Identifying Meaningful Citations. *AAAI Workshop: Scholarly Big Data*.

van Eck, N. J., & Waltman, L. (2010). Software survey: VOSviewer, a computer program for bibliometric mapping. *Scientometrics*, *84*(2), 523–538. https://doi.org/10.1007/s11192-009-0146-3

Whitley, R. (2000). *The Intellectual and Social Organization of the Sciences* (2nd edition). Oxford University Press.

Wilde, S. A., Zhao, G., & Sun, M. (2002). Development of the North China Craton During the Late Archaean and its Final Amalgamation at 1.8 Ga: Some Speculations on its Position Within a Global Palaeoproterozoic Supercontinent. *Gondwana Research*, *5*(1), 85–94. https://doi.org/10.1016/S1342-937X(05)70892-3

Wuchty, S., Jones, B. F., & Uzzi, B. (2007). The Increasing Dominance of Teams in Production of Knowledge. *Science*, *316*(5827), 1036–1039. https://doi.org/10.1126/science.1136099

Yang, H., Roeck, A. D., Gervasi, V., Willis, A., & Nuseibeh, B. (2012). Speculative requirements: Automatic detection of uncertainty in natural language requirements. *2012 20th IEEE International Requirements Engineering Conference (RE)*, 11–20. https://doi.org/10.1109/RE.2012.6345795

Zhai, M.-G., & Santosh, M. (2011). The early Precambrian odyssey of the North China Craton: A synoptic overview. *Gondwana Research*, *20*(1), 6–25. https://doi.org/10.1016/j.gr.2011.02.005

Zhao, G., Sun, M., Wilde, S. A., & Sanzhong, L. (2005). Late Archean to Paleoproterozoic evolution of the North China Craton: Key issues revisited. *Precambrian Research*, *136*(2), 177–202. https://doi.org/10.1016/j.precamres.2004.10.002

Zhao, G., Wilde, S. A., Cawood, P. A., & Sun, M. (2001). Archean blocks and their boundaries in the North China Craton: Lithological, geochemical, structural and P–T path constraints and tectonic evolution. *Precambrian Research*, *107*(1), 45–73. https://doi.org/10.1016/S0301-9268(00)00154-6

# Supporting Information

## S1 Text: Disciplinary artifacts in queries

The incidence of citation sentences returned by each query are not uniform across all signal and filter term combinations. There are far more publications from the Biomedical and Health Sciences (Bio & Health) than in other fields, accounting for a total of 47.5% of all publications indexed in the Web of Science Database; in contrast, publications in Math and Computer Sciences (Math & Comp) comprise a far smaller proportion of the database, accounting for only 3.1 percent.

Even accounting for the different number of publications per field, we still observe that some signal terms appear more in certain fields than expected, often as a result of differences in disciplinary jargon, topics, and norms (Figure SI 2b). For example, there are more *conflict** citances than expected in Social Sciences and Humanities (Soc & Hum), where it often appears in relation to conflict as a topic of study, such as the study of international conflict, conflict theory, or other interpersonal conflicts (Table SI 3, I). Similarly, *disprov** citances appear more often in Math & Comp, where disprove is often used in relation to proving or disproving theorems and other mathematical proofs (Table SI 3, II). Other notable differences are *controvers** citances appearing more often in Bio & Health, *debat** appearing most often in the Life and Earth Sciences (Life & Earth), and *disagree** appearing most in Physical Sciences and Engineering (Phys & Engr).

Filter terms are also non-randomly distributed across fields (Figure SI 2c). For example, the *+ideas* filter term appears more often than expected in Soc & Hum, possibly as a result of disciplinary norms around use and discussion of abstract theories and concepts (Table SI 3, III). In contrast, *+methods* is over-represented in Phys & Engr and Math & Comp, likely a result of these field's focus on methods and technique (Table SI 3, IV). Notably, *+studies* and *+results* are under-represented among Math & Comp publications, whereas *+ideas* and *+methods* are underrepresented among papers in the Biomedical and Health Sciences.

The complexity of disciplinary differences between queries is made apparent when examining combinations of signal and filter phrases (Figure SI 2d). While there are no obvious or consistent patterns between fields, there are particular differences by field. For example, compared to all other fields *contravers** citances are over-represented in Bio & Health (Table SI 3, V), except for *contravers +ideas,* which is instead slightly over-represented in Life & Earth. In contrast, *disagree** citances are under-represented in Bio & Health, but over-represented in Life & Earth and Phys & Engr (Table SI 3, VI). In some cases, the specific signal +filter term combination has a massive effect, such as *no consensus +ideas*, which is heavily over-represented in Soc & Hum (Table SI 3, VII), whereas all other signal and filter term combinations are under-represented. Similarly, *contradict* +ideas* and *contradict* +methods* are over-represented in Math & Comp (Table SI 3, VIII), whereas *+results* and *+studies* are underrepresented. Similar intricacies abound across the 325 combinations of signal term, filter term, and field, and demonstrate the importance that field plays in the utility and significance of our signal and filter terms.

Especially at the fine-grained field level, methodological artifacts can drive differences we observe between meso-fields. For example, in Soc & Hum, one of the meso-fields with the most disagreement was composed of papers from journals such as "*Political Studies"* and "*International Relations"*—journals and fields for which "debates" and "conflicts" are objects of study, which could confound the *debat** and *conflict** signal terms. This is demonstrated by the following invalid citation sentences:

1. "Since the late-1990s, there has been even less room for **debate** within the party (...)."
2. "Indeed, this whole idea harkens back to the badges of slavery of the 13th Amendment and the **debate** in (...)."
3. "In political behaviour literature, we refer to such **conflictive** opinions as "ambivalence" (...)."

4. “In politics as usual, people often do not like to see the **conflicts** and disagreements common to partisan debate (...).”

Even though terms such as “public debate”, and “parliamentary debate” were excluded (Table 1), the *debat** signal terms were over-represented in Soc & Hum (Figure SI 2); *conflict** was also overrepresented to a lesser extent. Interpretation of the results for main and meso-fields needs to be moderated by these, and other confounding artifacts (see the Supporting Information for further discussion).

## S2 Text: Robustness

To test the robustness of our results, we compare findings using the 23 queries with greater than 80% validity to those using the 36 queries with greater than 70% validity. The new queries include c*ontradict* _standalone_, contrary +studies, contrary +methods, conflict* +results, disagree* +methods, disagree* +ideas, disprov* +methods, disprov* +ideas, refut* +studies, refut* +results, refut* +ideas, debat* +ideas,* and *questionable +ideas.* Queries above the 80% validity cutoff account for 455,625 citances; the addition of 13 queries above the 70% cutoff bring this total to 574,020.

We find that our findings are robust whether using an 80% or 70% validity cutoff. Relaxing the validity cutoff results in including more citances, inflating the share of disagreement across all results. However, the qualitative interpretation of these results does not change (Table SI 4). The 80% and 70% cutoffs both produce the same ordering of fields from most to least disagreement. Similarly, the ordering of fields from high-to-low disagreement holds between the 80% and 70% cutoff for all quantities presented here, including the average change per year, the ratio of disagreement between non-self-citation and self-citation, and the average change in disagreement per age bin. Some fields gain more from these new queries more than others, manifesting in more or less intense field differences in findings. For example, Soc & Hum gains a full 17 percentage points in overall disagreement with the 70% threshold, with the increase across all fields at only 8 points. Similarly, the ratio of non-self-citation to self-citation is 2.2x for Math & Comp with the 80% cutoff, but only 1.3x for the 70% cutoff. This likely stems from the relative skew of the added queries to certain fields, leading to larger gains in disagreement.

## S3 Text: Disagreement by contextual factors

### Paper age

Other contextual factors may relate to disagreement. For example, authors may disagree with more recent papers at different rates than older ones. We quantify disagreement based on the age of a cited paper, relative to the citing paper. Following a brief *bump,* or increase in disagreement (at 05-09 years), older papers tend to be receive fewer disagreement citances (Figure SI 4), a pattern driven by field differences. Low consensus, high complexity fields such as Soc & Hum and Bio & Health both exhibit a clear decreasing pattern, with falling disagreement as the paper ages. Life & Earth, in the middle of the hierarchy, repeats this pattern, but only after a period of stability in disagreement in the first ten years. Disagreement instead steadily increases over time in high consensus and low complexity fields such as Phys & Engr and Math & Comp.

### Position in the paper

Disagreement is not equally likely to occur throughout a paper. Investigating the distribution of disagreement citances across papers, we find that they are far more likely to occur in the beginning of a paper, likely in the introduction, and then towards the end, likely the discussion section (Figure SI 5), corresponding to previous observations of disagreement cue phrases in PLoS journals (Bertin et a., 2016), and likely indicating a unique argumentative role of disagreement. The precise patterns differ by field. For example, in *Soc & Hum*, disagreement citances are more evenly distributed through the first 40 percent of the paper, whereas in *Bio & Health* and *Life & Earth* disagreement citations are more likely to appear near the end of a paper. While these field level differences may reflect differences in how fields use citations, they are more likely the result of distinct article formats across fields (i.e., long literature reviews in *Soc & Hum*).

### Gender of citing-paper author

Men and women authors may be more likely disagree at different rates. To test this, we infer a gender for the first and last authors of papers with a disagreement citance published after 2008, determined based on the author's first name as in past work (Larivière et al., 2013). Overall, there is little difference in the rate of disagreement between men and women first and last authors (Figure SI 6). The one exception is Math and Computer science, where women last authors disagree 1.2 times more often than men, though the rate of disagreement is small, and driven by a small number of instances.

### Self-citation

We investigate the extent to which disagreement differs based on self-citation, that is when there is in the presence of overlap between the authors in a citing and the cited papers. We would expect that authors will be less likely to cite their own work in the context of disagreement, which is affirmed by our indicator. Overall, the rate of disagreement among non-self-citing papers is 2.4 times greater than for self-citation citances (Figure SI 3). The field with the largest difference is Bio & Health (2.5 times greater), followed by Phys & Engr (2.2 times greater), Math & Comp (2.2 times greater), Life & Earth (1.9 times greater), and finally, Soc & Hum (1.6 times greater).

## S4 Text: Papers with most disagreement citances

While the extensive manual validation of our queries and results ensures the robustness of our analysis at an aggregate scale, the list of publications issuing most disagreement citations does reveal that it remains difficult to separate research object from commentary on cited material. Table SI 6 shows the papers that issue the most disagreement citations; going through these papers reveals several artifacts. "Debating" and "debates" are, for example, the object of study in Alén et al. (2015) and Doody & Condon (2012), and the citances in the paper reflect this, e.g. "students also seem to both enjoy debates and recognise their value" and "debate is effective in helping students learn a discipline and demonstrate the ability to read and write critically." Controversy, likewise, is the subject in Ersoy (2010), Nam (2014), and Colston & Vadjunec (2015), as are environmental conflicts in Stepanova & Bruckmeier (2013). Bruschke & Divine (2017) makes frequent mention of the first televised US presidential *debate*. This leaves French & Koeberl (2010), Kalter (2003) and Millan (2006), three publications that do not immediately appear focused on subjects that would trigger our queries.

French & Koeberl (2010) discusses methods for identifying meteorite impact structures on earth, "as well as an overview of doubtful criteria or ambiguous lines of evidence that have erroneously been applied in the past", and this paper indeed cites many sources in the context of controversy, e.g. "the identification of such glasses as impact or non-impact products is difficult and commonly controversial," "the impact origin for many glasses still remains controversial and unconfirmed," "there are also debates about the formation of maskelynite itself" and "the nature, characteristics, and causes of these changes have been widely studied and are still being debated."

Kalter (2003), while classified as a full-length article, is a book that was also included in a special issue of the journal Neurotoxicology and Teratology. Considered a pillar in the study of congenital abnormalities, its exceptional length may account in part for its high number of disagreement citances. One of these citances describes "a discussion—debate better characterizes it—that took place in 1953," but many others indeed refer to scientific disagreements within the field of study, e.g. "a Mayo Clinic study seemed to agree, despite conflicting evidence," "a contrary finding came from Scotland, another high-risk region," "an early analysis, as well as a later one, disagreed" and "earlier retrospective and prospective studies had been contradictory."

Millan (2006) is a review article making a case for multi-target agents to treat depressive states. It likewise introduces a number of citances that indeed signify disagreement in the scientific literature, e.g. “the gravity of cognitive impairment in young patients is still debated,” “this notion remains somewhat controversial,” “its precise degree of efficacy in this regard is still debated” and “for recent critical discussions of these controversial issues–from a variety of viewpoints–see […].” However, a few false positives also occur, when citing the work of an author by the name of DeBattisa, whose name was caught by our *debate* query.

## S5 Text: papers most often cited in the context of disagreement

We also examine disagreement form the cited paper perspective, that is, by looking at those papers that received the most paper-level disagreement citations and which were most often cited in the context of community-level disagreement. Table SI 7 lists these papers, and again we reveal issues with methodological artifacts, but also highlight interesting instances of controversy in the literature. The majority of these publications relate to plate tectonics, and in particular, the North China Craton. These papers include Zhao, Sun, Wilde, and Sanzhong (2005), Kusky and Li (2003), Zhao, Wilde, Cawood, and Sun (2001), Zhai and Santosh (2011)¸ Wilde, Zhao, and Sun (2002) and Kusky (2011). Li et al. (2008) also appears to be closely related. Several of these publications have authors in common, notably Zhao (3), Wilde (3), Sun (3), Kusky (2) and Li (2). Zhai and Santosh (2011) summarizes the situation as follows:

> “A long controversy and debate surround the evolution of the NCC, particularly the timing and tectonic processes involved in the amalgamation of the Eastern (Yanliao) and Western (Ordos and Yinshan) Blocks along the Central Orogenic Belt. One school of thought proposes an east-directed subduction of an old ocean, with final collision between the two blocks at ~ 1.85 Ga <several citations to Zhao’s work>. In contrast, some others suggest a westward subduction, with final collision between the two blocks to form the NCC at ~ 2.5 Ga <several citations to Kusky’s work>.”

The majority of disagreement citances to these papers also mention this long-standing scientific controversy, including phrases such as,

- “the number of continental blocks and the mechanism by which they were welded together to form the coherent basement remain controversial,”
- “tectonic history of this central region is in debate,”
- “it is currently debated how the collisional processes proceeded,”
- “controversy still remains as to the timing and tectonic processes involved,”
- “it still remains controversial as to how the craton should be subdivided and where the collisional boundaries are located,”
- “models that evaluate the Paleoproterozoic crustal evolution of the NCC remain controversial,”
- “the timing of the collision between these blocks remains controversial,”
- “this controversy is also reflected in various tectonic models for the NCC" and
- “the time of the amalgamation between the Eastern Block and the Western block is still debated,”

These phrases are often accompanied by a large number of citations to papers by Zhao and Kusky, with both authors either explicitly posed as opposing one another like in the example from Zhai & Santosh (2011) above, or citations to their papers grouped together as if to present a large body of conflicting literature. It is clear that this scientific controversy dominates scholarship on the North China Cranton, to the point that even Wikipedia mentions it extensively (“North China Craton,” 2020). It should be noted that the works by Zhao and Kusky also receive a fair share of citations that mention their empirical and theoretical contributions, without the context of disagreement, making material research contributions to both their preferred model and the research on the cranton at large. The high share of disagreement citances to these papers appears to stem from the highly divided nature of their research field.

The three remaining papers cover different topics. Munro (2003) reviews literature on lipids in cell membranes of eukaryotic cells and discusses the (non)existence of lipid rafts. It points out that, regardless of enthusiasm for the lipid raft model in the research community, observations of these raft structures are problematic and several factors exist that cast doubt on the model. The existence of lipid rafts should therefore be treated as hypothetical rather than established fact. Citances that mention this paper in the context of disagreement also appear to use it as an exemplar of controversy in the field, with phrasing such as,

- "the entity of lipid microdomains is controversial,"
- "the existence of noncaveolar lipid rafts in vivo is still debated,"
- "this early operative definition of lipid rafts was subject of much debate,"
- "the lipid raft concept has been controversial since it was introduced several decades ago" and
- "a number of points have been recently debated in the literature."

Even many citances that do not qualify as disagreement per our operationalization appear to still signal it, e.g.

- "rafts still remain a hypothesis,"
- "evidence that lipid rafts exist in living cells remains elusive,"
- "the classical perception of rafts as stable entities within the fluid bilayer has provoked some opposition,"
- "the biological substrate for this notion is not clearly defined" and
- "isolation of DRM or lipid rafts is however a delicate matter."

This paper, specifically, serves as a focal point of controversy within the community researching lipid rafts precisely because its primary purpose appears to be to create this controversy; raising concerns about the lipid raft model and calling for a reevaluation of its canonicity in the face of shaky foundational evidence of the existence of these rafts. This is different from the North China Cranton papers, where controversy is long established in the field.

Murphy et al. (2009) focuses on a pre-processing method used in low-frequency fMRI research, called global signal regression. This paper alleges that the method is inadequate and "may cause spurious findings of negatively correlated regions in the brain." This paper appears to have heralded a change in how data is handled in this research field, with many citances mentioning it to explain why data was processed differently, e.g.

- "the global signal was not regressed out due to its controversial biological interpretations,"
- "given the controversy of removing the global signal in the preprocessed rs-fMRI data, we did not regress the global signal out in the present study,"
- "global signal regression is a somewhat controversial part of the preprocessing pipeline for resting state MRI data, and was not performed in this study" and
- "given the controversy of removing the global signal in the preprocessing of R-fMRI data […], we did not regress the global signal out."

Other citances in the context of disagreement simply point out this controversy, e.g.

- “there is ongoing debate as to the nature of anti-correlations introduced by global signal regression”
- “in recent years there has been an ongoing debate on global signal removal in the preprocessing.”

As with the lipid raft paper, many citances that do not qualify as disagreement also embrace the controversy, as evident in phrasing such as “global-signal was not included in the model for its effects on brain–behavior correlations” and “we decided against this approach as several recent studies showed that global signal regression may significantly bias connectivity analyses.”

Finally, Debat et al. (2003) is another example of a false positive result of our approach, in which the lead author’s name activated the *debate* query. While this is unfortunate, our extensive manual validation of our query results shows that despite this prominent false positive, there were no large systemic flaws in our approach that might otherwise color our analysis at the aggregate level. While eliminating cited author names from citances at scale is not trivial, this example serves to stress the importance of text pre-processing.

## S6 Text: Disagreement and citation impact

To address whether publications cited in the context of disagreement citation were cited at different rates than their counterparts, we compared the number of citations received in year *t+1* for papers that featured in a disagreement citance for the first time in year *t*, with the average number of citations received in year *t+1* by papers that received the exact same number of citations in year *t*. This over- or undercitation of individual papers that encountered disagreement can then be aggregated to arrive at the average under- or overcitation of ‘disagreed-with’ papers in general.

We define *t* as the time in years since publication and *c* as the number of citations a paper received at time *t*. We calculate for each combination of *t* and *c* the number of papers $p_{c,t}$ that were first cited in the context of disagreement at time *t* when they held *c* citations. Using these, we calculate the number of citations received by these papers in the year following publication, averaged across all combinations of *t* and *c*,

$$\bar{c}_{next,disagreement} = \frac{\sum_c \sum_t p_{c,t} * \bar{c}_{next,disagreement,c,t}}{\sum_c \sum_t p_{c,t}}$$

In the same way, we also calculate the expected number of citations, defined using the average number of citations received by papers that received *c* citations in year *t*, regardless of whether they were cited by a disagreement citation.

$$\bar{c}_{next,expected} = \frac{\sum_c \sum_t p_{c,t} * \bar{c}_{next,expected,c,t}}{\sum_c \sum_t p_{c,t}}$$

We calculated *d* as the ratio of these two values. When greater than one, it indicates that papers received more citations than expected in the year after having been cited in the context of disagreement. A value less than one indicates that papers with a disagreement citation received fewer citations in the year following.

$$d = \frac{\bar{c}_{next,disagreement}}{\bar{c}_{next,expected}}$$

The results of this analysis (Table SI 5) show that being cited in a context of disagreement has little to no effect on the citations received by papers in the year following their citation (or not) in the context of disagreement. Extending the analysis to citations received in year *t+2* and *t+3* yielded similar results.

We also investigate whether disagreement relates to the number of citations a paper receives. First, we examine the citing paper perspective, identifying the 3.5 percent (n = 126,250) of publications that contain at least one disagreement citance in their text. Across all publications, those with at least one disagreement citance tended to receive more citations than those without disagreement in the first four years, beginning with one additional citation in the first year following publication, and expanding to a difference of about 4.7 citations by the fourth year (Figure SI 7), an effect that varies, yet is qualitatively consistent across all fields. This effect may be confounded by article type—for example, review articles are over-represented in terms of disagreement—24.6 percent of all review articles contain a disagreement citance—and review articles are also known to be more highly cited (Miranda & Garcia-Carpintero, 2018). While excluding review articles does shrink this gap, the citation count for full research articles (85 percent of all publications) remains 2.5 citations higher for those with a disagreement than for those without.

We note that these results are confounded by our umbrella definition of disagreement, which does not differentiate between *paper-level* and *community-level* disagreement. Paper-level disagreement, that is when the author of the citing paper explicitly contrasts their study with another, is a straightforward example of issuing (by the citing paper) and receiving (by the cited paper) disagreement. Community-level disagreement, in contrast, either involves a citing author rhetorically positioning two or more papers as being in disagreement, or citing a past paper, such as a review, as evidence of the controversy surrounding a topic. While these two cases offer evidence of disagreement in the field, their potential for identifying specific, controversial papers, as in Radicchi (2012), is less clear. Future research should aim to disentangle *paper-level* and *community-level* disagreement, and understand their varying relationship to citation impact.

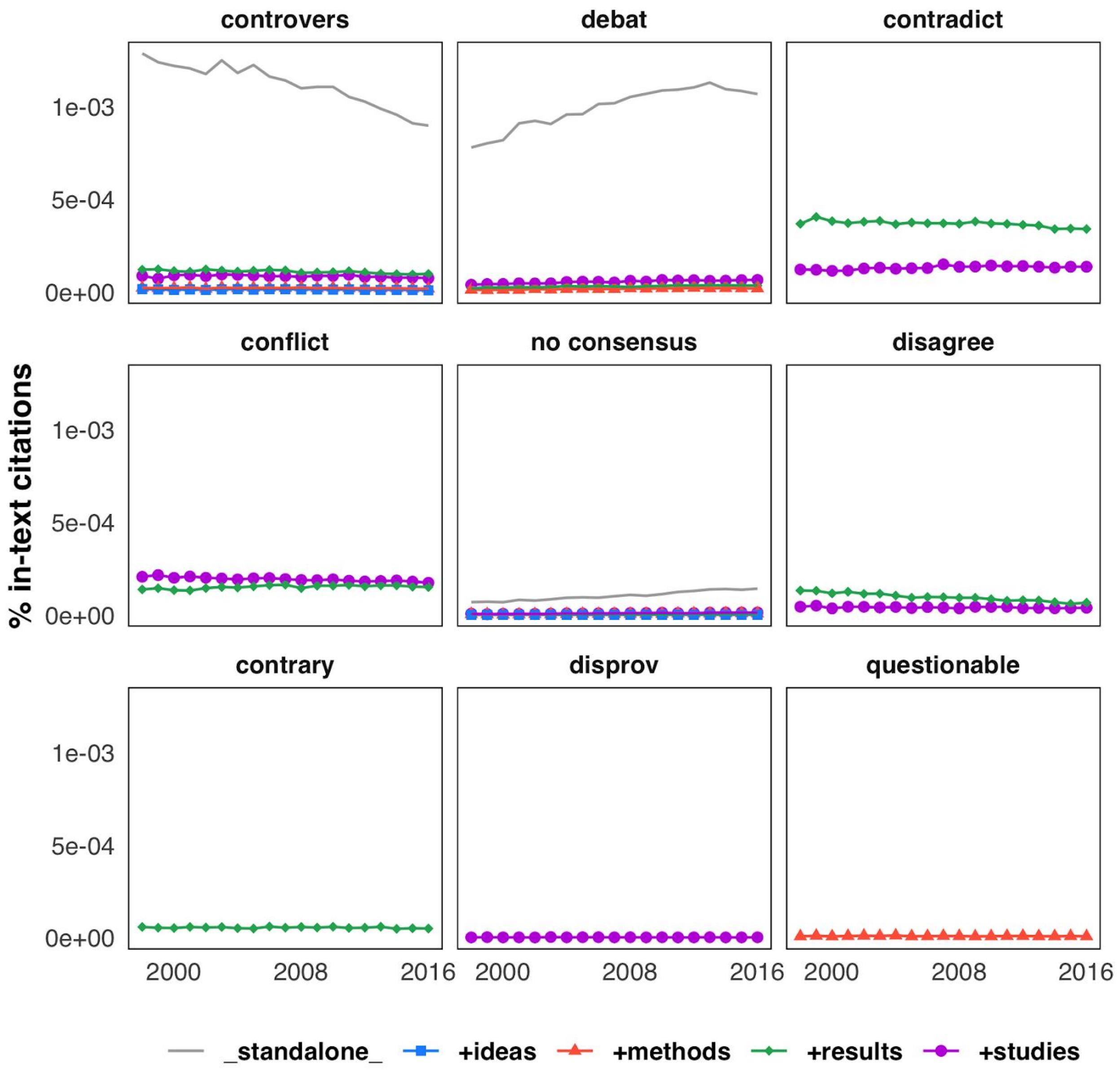

***Figure SI 1.*** *Percent of all citances returned by each of the 23 queries with validity over 80 percent. Each panel corresponds to the signal phrase, and lines within each panel to filter phrases.*

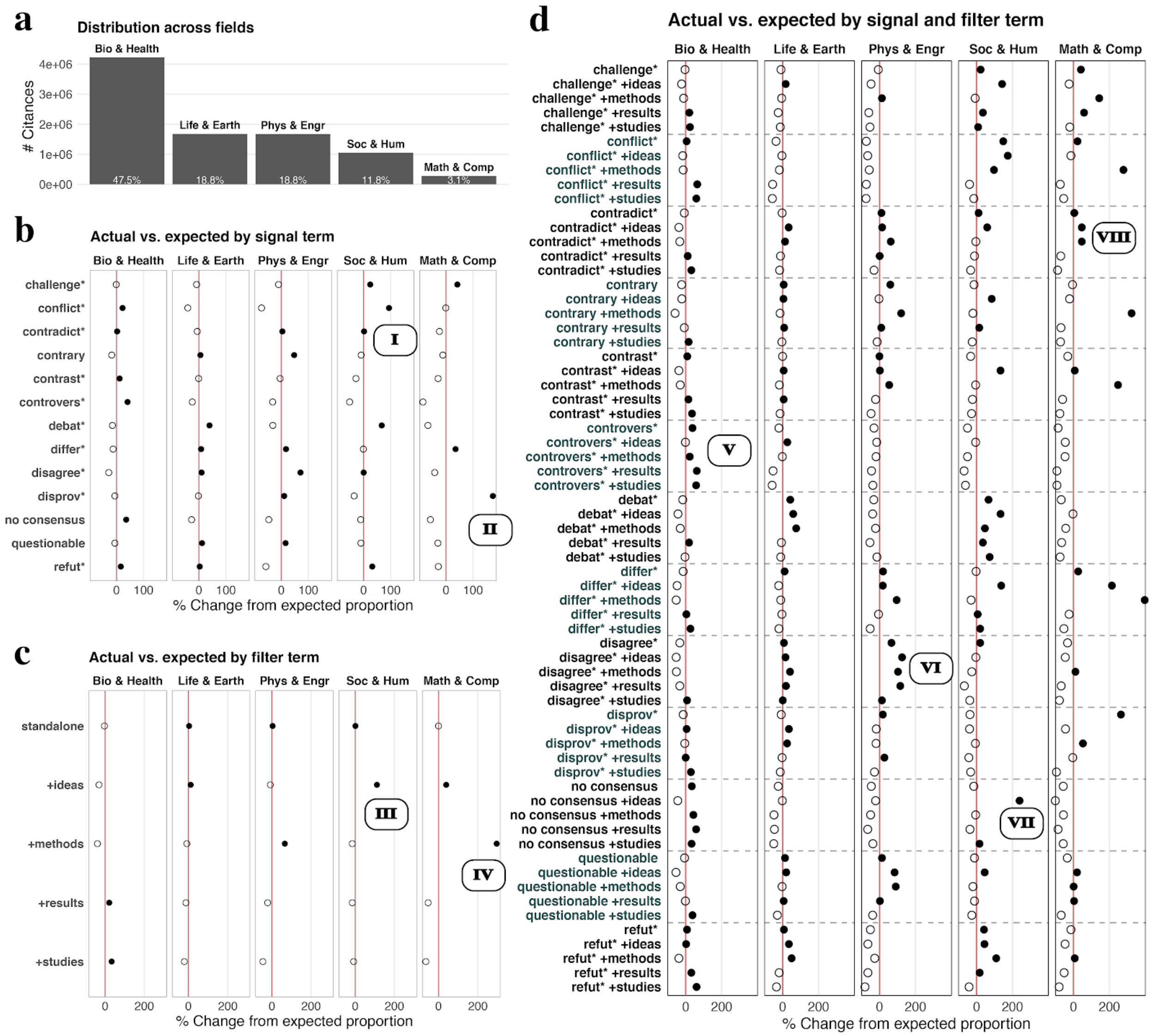

***Figure SI 2.*** *Distribution of citances returned by signal/filter term queries. Callouts (I, II, …, VIII) map to examples in Table SI 3. a. Distribution of all disagreement citances appearing in papers across five fields: Biomedical and Health Sciences, Life and Earth sciences, Physical Sciences and Engineering, Social Sciences and Humanities, and Math and Computer Science. b-d. Percentage change between the actual number of citances per field and signal/filter term combination compared to the expected given the disciplinary distribution (from a). The red line corresponds to 0 percent increase between the actual and expected. White dots indicate that the citances for that signal/filter term are under-represented (lower than expected, ratio less than zero), whereas black dots indicate that citances are over-represented (more than expected). Shown aggregated across signal terms (b), filter terms (c), and for all signal/filter term combinations (d).*

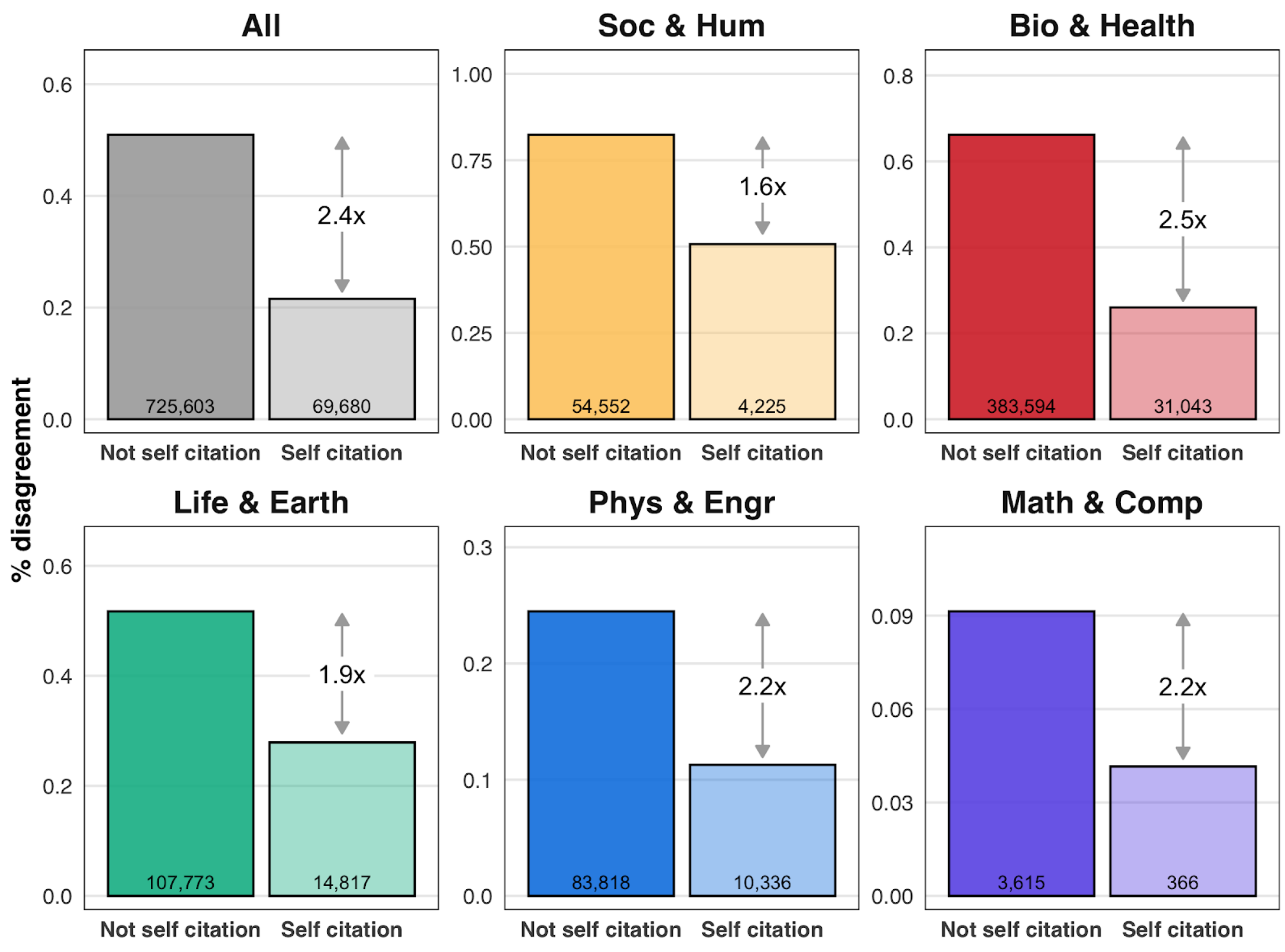

***Figure SI 3***. *Authors disagree less when citing their own work. Percentage of disagreement citances among instances of non-self and self-citation, 2000-2015. A citance is defined as a self-citation when the citing and cited paper have at least one name in common. Results are shown by field. Numbers below each bar are the number of disagreement citances. Overall, disagreement is 2.4 times more common for non-self citation than for self-citation, with variance between major fields.*

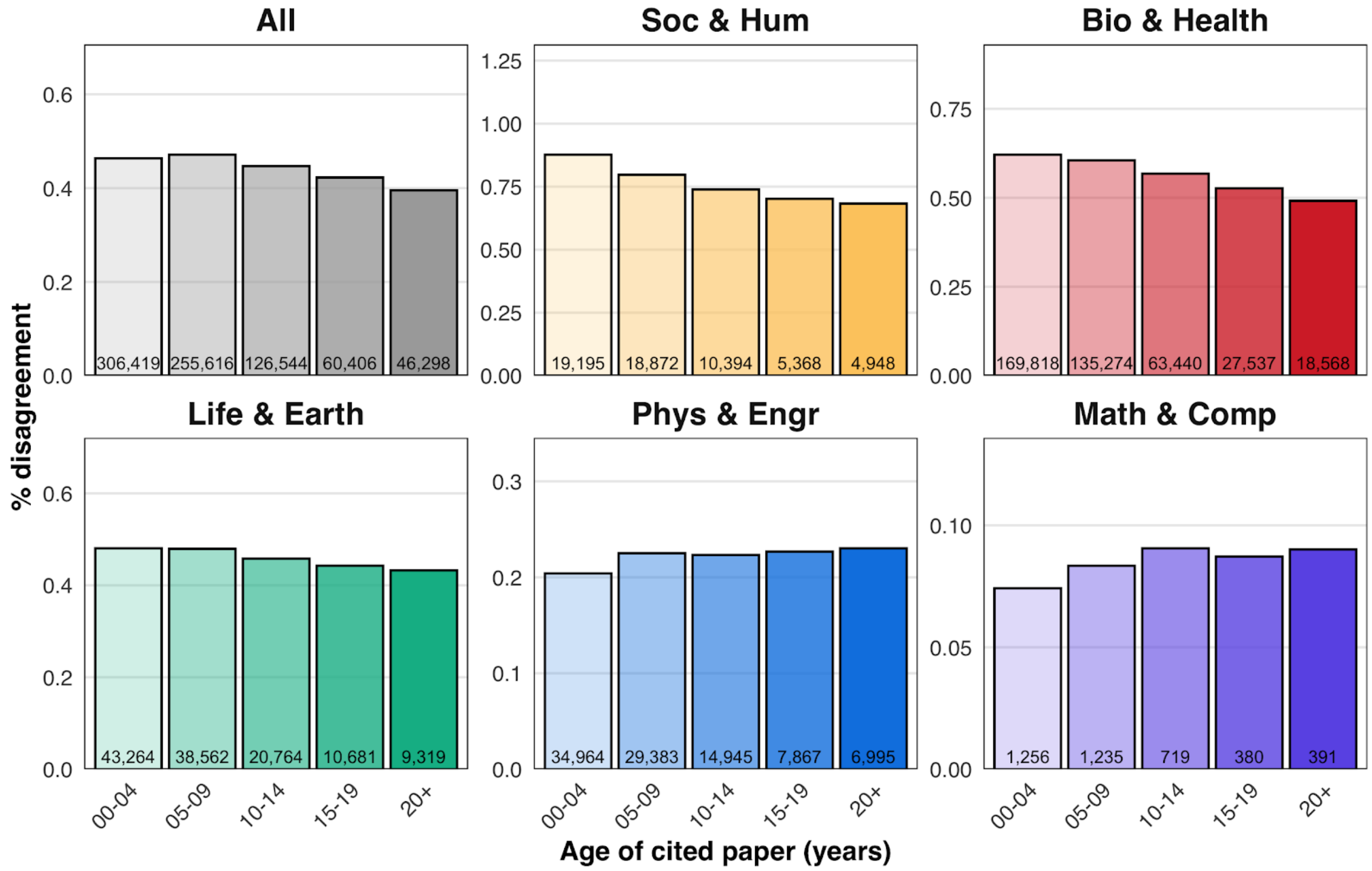

***Figure SI 4.*** *On average, older papers are less likely to receive a disagreement citance, though this trend does not hold for the "hard" sciences. Percentage of disagreement citances by the relative age of the citing to the cited paper, in years, and high-level field, for papers published between 2000 and 2015. Intensity of color corresponds to the age category of the cited paper.*

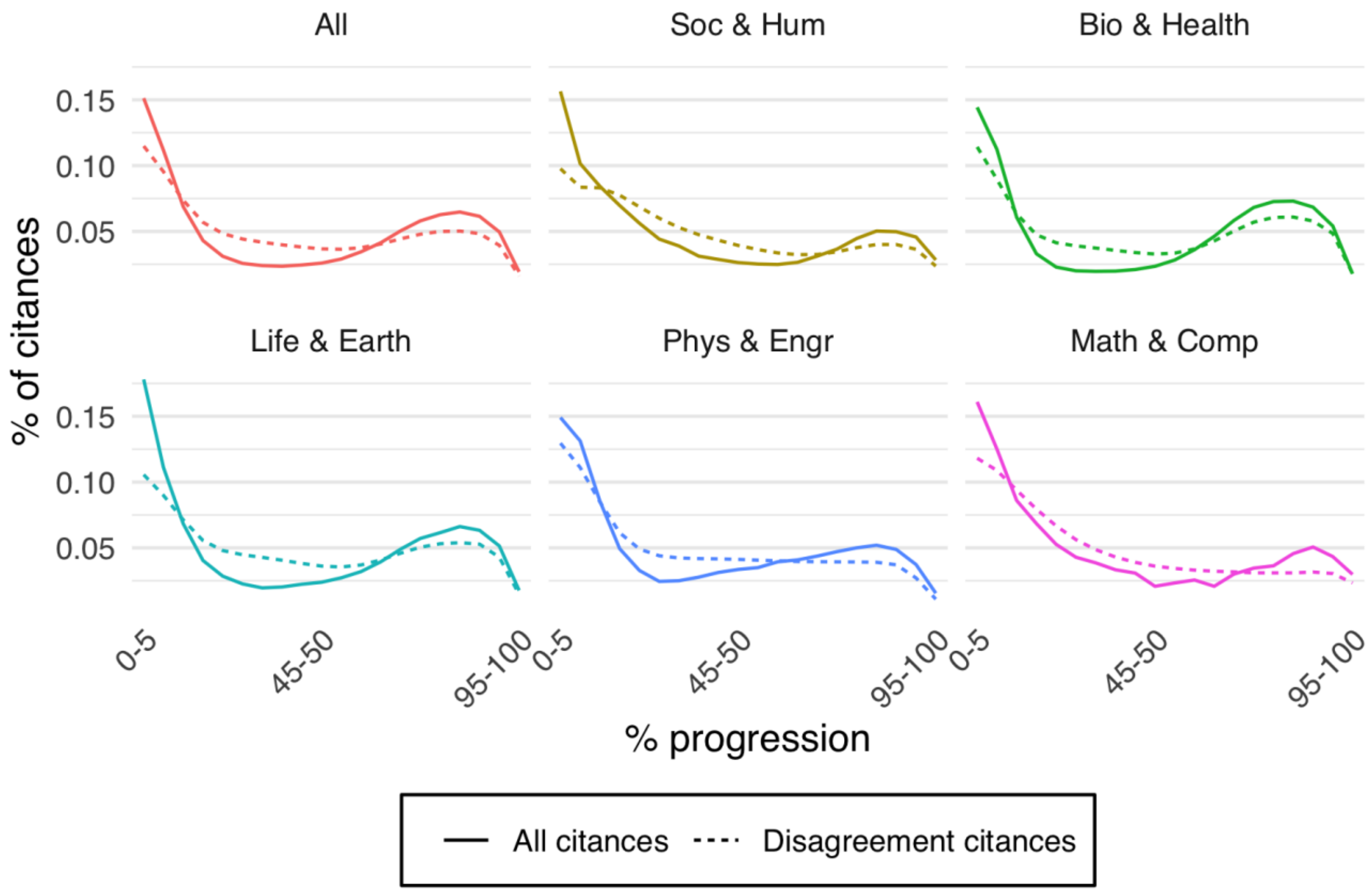

***Figure SI 5***. *Distribution of citances by their position in the text of the manuscript, and by field. Shown for all citances (solid line) and disagreement citances (dotted line). For example, about 15% of disagreement citances in Physical Sciences and Engineering appear in the first 0-5 percent of the sentences in documents.*

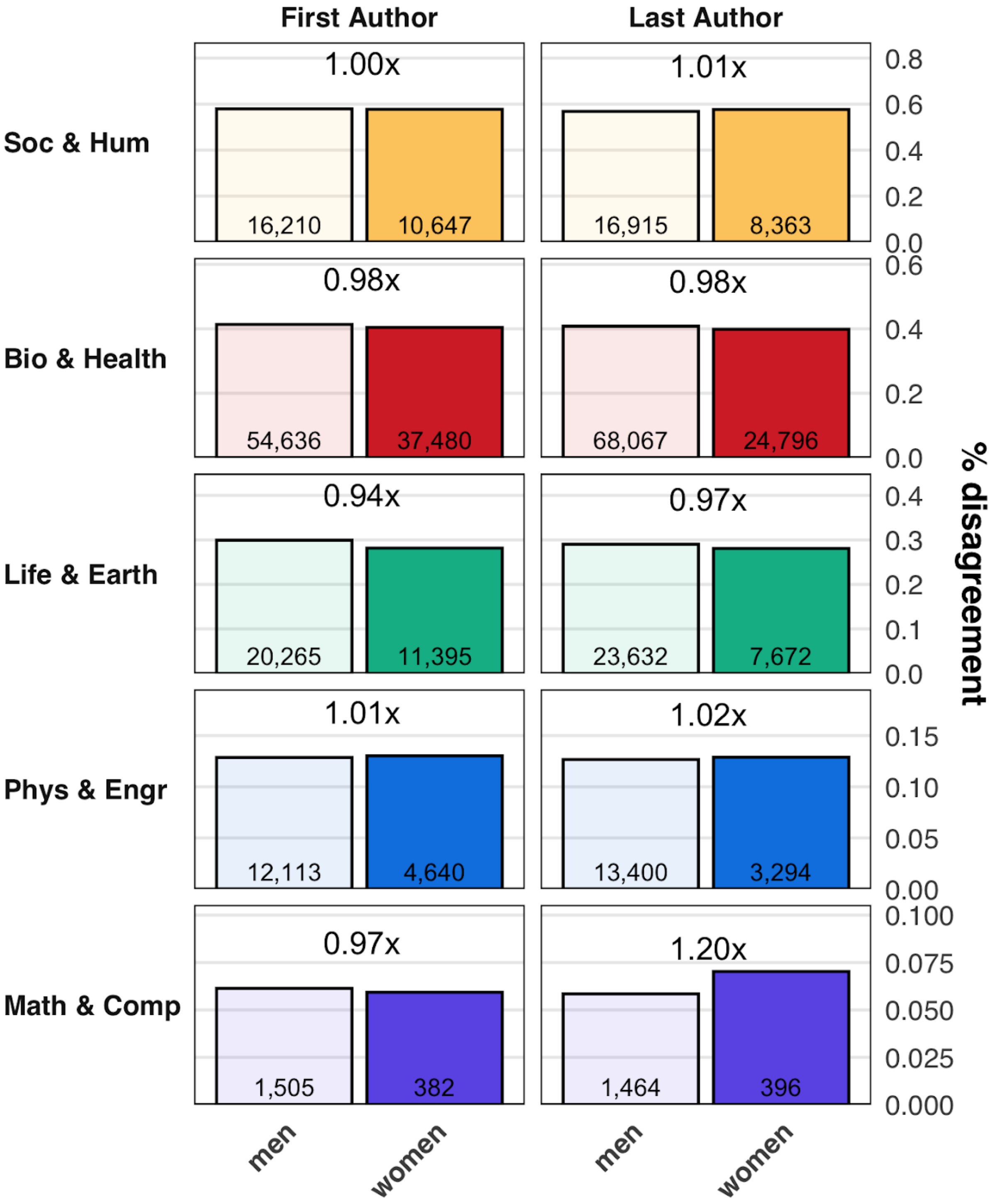

***Figure SI 6.*** *Little difference in disagreement between men and women.* *Percentage of disagreement citances by gender of the citing-paper author, their authorship position (first or last), and the high-level field. Numbers above each bar corresponds to the ratio difference between the percentage of disagreement between women and men. The number below each bar corresponds to the number of disagreement citances. we infer a gender for the first and last authors of papers with a disagreement citance published after 2008, determined based on the author's first name as in past work (Larivière et al., 2013).*

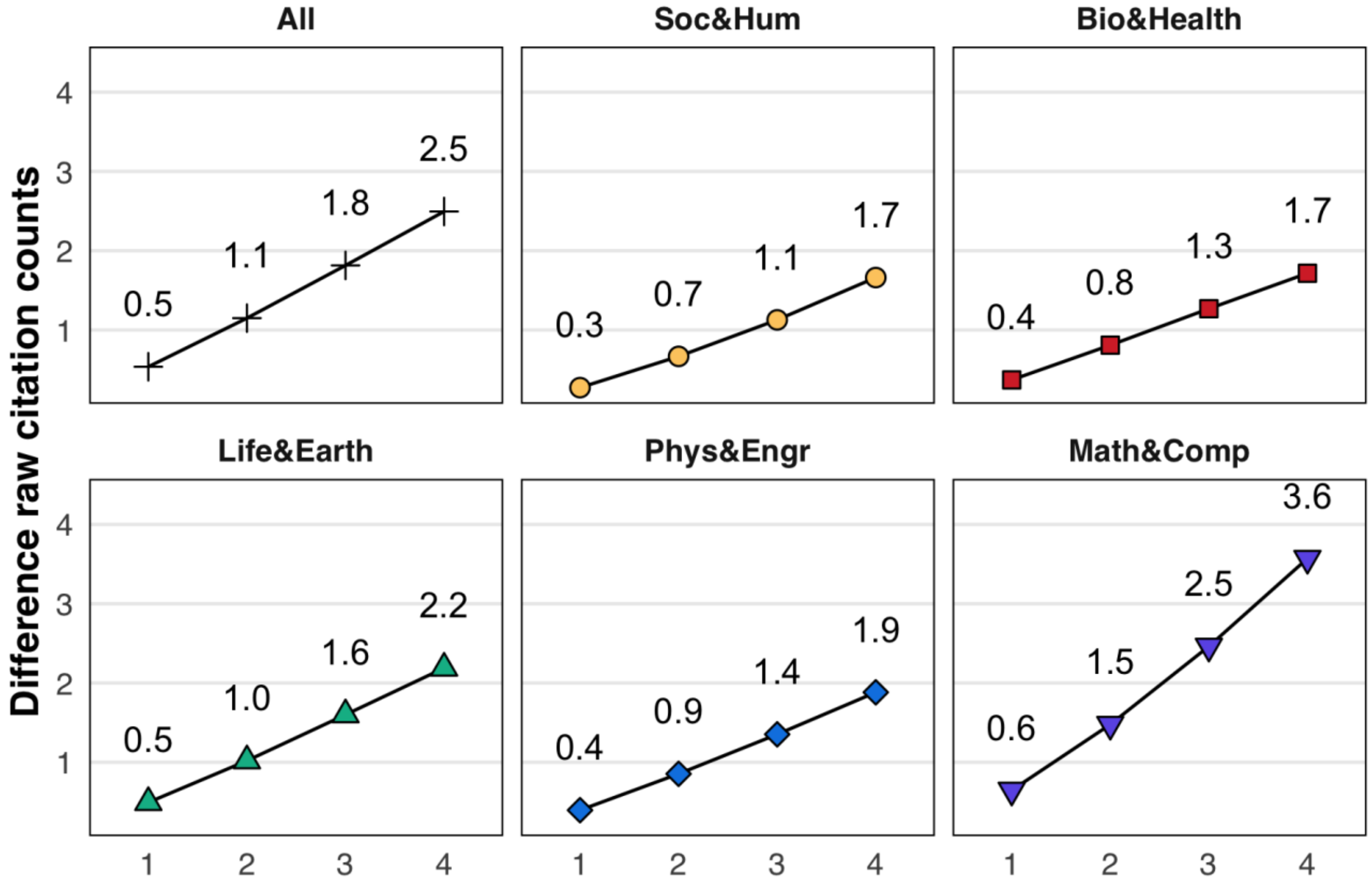

***Figure SI 7.*** *Full research articles with a disagreement citance are cited more. The y-axis shows the difference in average citation counts for papers containing at least one disagreement citance, and for papers without. Positive values indicate that publications with disagreement received more citations than those without. Values are shown for the population of publications in each year following publication (x-axis). Shown here for only articles labeled in the Web of Science database as full research articles.*

***Table SI 1.*** *Number of citances in the Elsevier ScienceDirect database containing signal term (rows) and filter term (columns) combination. For some signal terms, variants are excluded; for example, "not contradict" is not matched. For filter terms, "_standalone_" indicates that only the signal term was used to query. The remaining columns, +studies, +ideas, +methods, and +results correspond to sets of filter terms outlined in Table SI 2.*

| | **_standalone_** | **+studies** | **+ideas** | **+methods** | **+results** |
|---|---|---|---|---|---|
| ***challenge**** | 405,613 | 16,120 | 7,114 | 13,806 | 15,352 |
| ***conflict**** | 212,246 | 22,190 | 3,603 | 5,560 | 49,961 |
| ***contradict**** | 115,375 | 19,509 | 5,482 | 2,793 | 52,648 |
| ***contrary*** | 171,711 | 17,207 | 3,651 | 4,699 | 27,273 |
| ***contrast**** | 1,257,866 | 116,450 | 7,774 | 37,372 | 119,181 |
| ***controvers**** | 154,608 | 12,187 | 1,840 | 3,028 | 15,473 |
| ***debat**** | 150,617 | 8,509 | 1,774 | 2,678 | 4,663 |
| ***differ**** | 2,003,677 | 100,764 | 9,531 | 85,309 | 110,599 |
| ***disagree**** | 52,615 | 5,724 | 1,142 | 1,682 | 12,459 |
| ***disprov**** | 2,938 | 278 | 528 | 100 | 358 |
| ***no consensus*** | 16,632 | 1,424 | 37 | 830 | 421 |
| ***questionable*** | 24,244 | 1,045 | 852 | 1,175 | 2,050 |
| ***refut**** | 10,322 | 1,399 | 1,564 | 338 | 2,262 |

***Table SI 2.*** *Examples of valid and invalid citances returned for "challenge*" and filter term combinations. Signal terms are bolded and underlined, whereas relevant filter terms are underlined but not bolded.*

| Filter term | Valid | Sentence | Reference |
|---|---|---|---|
| **_standalone_** | Yes | Although phosphorus has traditionally been seen as the limiting nutrient in freshwater ecosystems (e.g. Blake et al., 1997; Karl, 2000), more recent evidence has begun to **challenge** this view and has demonstrated that both nitrogen and phosphorus can limit, or at least co-limit, primary production in freshwaters (e.g. Elser et al., 2007; Francoeur, 2001). | Smith et al., 2009 |
| **_standalone_** | No | Analogs of these molecules have shown up to 1000-fold higher activity but are a great **challenge** to delivery because of their extreme hydrophobicity [33]. | Brannon-Peppas & Blanchette, 2004 |
| **+studies** | Yes | However, recent studies have **challenged** this survival benefit in comparison with current usual care.4-6 | Gottlieb et al., 2015 |
| **+studies** | No | The low affinity with which volatile general anesthetics bind to macromolecules has made conclusive identification of the in vivo targets by direct binding studies a **challenge** [1]. | Zhang & Johansson, 2005 |
| **+ideas** | Yes | This result **challenges** AUM theory (Gudykunst, 2005) and some prior research (e.g. Gao & Gudykunst, 1990). | Rui & Wang, 2015 |
| **+ideas** | No | The description of the resonant electron capture by molecules connected with the formation of negative ions represents still the **challenge** for the **theory** [1]. | Papp et al., 2013 |
| **+methods** | Yes | This model has since been **challenged** by claims that Helderberg formation boundaries are isochronous across the basin (Anderson et al., 1984; Demicco and Smith, 2009). | Husson et al., 2015 |
| **+methods** | No | Subsequent studies in the human **challenge** model have also supported the role of NA-specific antibody in protection. [34] | Treanor, 2015 |
| **+results** | Yes | It has been reported that the prevalence of autoimmune disorders in celiac disease is related to the duration of exposure to gluten [34], although this result has been **challenged** [35]. | Skovbjerg et al., 2004 |
| **+results** | No | Some of the larger **challenges** identified in Africa include data collection, access and management, infrastructure and capacity (Han et al., 2014). | Stephenson et al., 2017 |

***Table SI 3.*** *Examples of citances from notable signal/filter phrase combinations labeled in Figure XYZ. "[...]" has been used in places of reference names or numbers. Relevant signal terms have been bolded and underlined, whereas relevant filter terms have only been underlined.*

| Label | Signal | Filter | Valid | Example |
|---|---|---|---|---|
| **I** | conflict* | None | No | "For instance, […] study **conflicts** based on ethnicity where ethnic identity works as a device to enforce coalition membership." |
| **II** | disprove* | None | No | "These techniques are typically used to confirm or **disprove** an a priori hypothesized model, i.e. to test the statistical adequacy of a proposed causal model […]" |
| **III** | challenge* | +ideas | Yes | "Reversal theory **challenges** the idea of personality traits in suggesting that people fluctuate between metamotivational states that are opposite and mutually exclusive […]" |
| **IV** | contradict* | +methods | Yes | "The existence of topological singularities is in **contradiction** with […]'s method of continuous transformations of a rectangular Cartesian frame of coordinates into a curvilinear grid without singularities […]." |
| **V** | controvers* | +ideas | Yes | "There is still a considerable **controversy** about the idea of homology between specific areas of rat and primate PFC […]." |
| **VI** | disagree* | +results | Yes | "These results were in **disagreement** with much of the literature, where there is consensus that, H2 is almost exclusively catalyzed by SRB at low COD/SO42- ratios […]" |
| **VII** | no consensus | +ideas | Yes | "Because of the controversial data collected, **no consensus** about this theory has been reached to date […]." |
| **VIII** | contradict* | +methods | Yes | "Thus, the approach is somewhat **contradictory** to certain approaches in robotics that strive to develop highly autonomous robots capable of performing independent decisions based on sensory data […]." |

***Table SI 4.*** *Results are robust to both the 80% and 70% validity cutoffs. Quantities of interest using the 23 queries above the 80% validity cutoff, and the 36 queries above the 70% validity cutoff. Shown are the overall rates of disagreement and the change in the share of disagreement per year.*

| **Quantity** | **Validity cutoff** | **All Fields** | **Soc & Hum** | **Bio & Health** | **Life & Earth** | **Phys & Engr** | **Math & Comp** |
|---|---|---|---|---|---|---|---|
| **Overall** | 80% | 0.32% | 0.61% | 0.41% | 0.29% | 0.15% | 0.06% |
| | 70% | 0.40% | 0.78% | 0.50% | 0.36% | 0.20% | 0.15% |
| **Change per year** | 80% | -0.0005 | -0.0033 | +0.0017 | +0.0018 | -0.0045 | -0.0019 |
| | 70% | -0.0005 | -0.0042 | +0.0022 | +0.0019 | -0.0061 | -0.0028 |

***Table SI 5.*** *Being cited in the context of disagreement has little impact on citations in the year following. For each field, shown are the number of cited papers, as well as for t+1, t+2 and t+3 with t being the year in which a cited paper first featured in the context of disagreement, its average number of received citations, expected number of received citations, and d the ratio between these two values. When d is greater than one, papers cited in the context of disagreement receive more citations in the following year than expected. When d is less than one, they receive fewer citations than expected.*

| Scientific field | # Records | Avg. citations, t+1 following disagreement | Expected citations, t+1 following disagreement | $d_{t+1}$ | Avg. citations, t+2 | Expected citations, t+2 | $d_{t+2}$ | Avg. citations, t+3 | Expected citations, t+3 | $d_{t+3}$ |
|---|---|---|---|---|---|---|---|---|---|---|
| All | 109,545 | 3.03 | 3.08 | 0.983 | 3.02 | 3.05 | 0.990 | 2.96 | 2.98 | 0.993 |
| Bio & Health | 60,707 | 2.73 | 2.81 | 0.969 | 2.68 | 2.75 | 0.974 | 2.56 | 2.65 | 0.966 |
| Life & Earth | 20,581 | 3.43 | 3.35 | 1.023 | 3.55 | 3.42 | 1.038 | 3.63 | 3.44 | 1.056 |
| Math & Comp | 770 | 3.36 | 3.34 | 1.005 | 3.54 | 3.28 | 1.080 | 3.29 | 2.97 | 1.109 |
| Phys & Engr | 18,011 | 3.55 | 3.52 | 1.006 | 3.48 | 3.44 | 1.010 | 3.43 | 3.34 | 1.027 |
| Soc & Hum | 9,476 | 3.04 | 3.11 | 0.979 | 3.20 | 3.28 | 0.975 | 3.30 | 3.40 | 0.971 |

*Table SI 6. Papers that introduced the most disagreement citances*

| Total citances | Disagreement citances | Publication | Title | Document type |
|---|---|---|---|---|
| 50 | 27 | Alén, Domínguez, & De Carlos (2015) | University students' perceptions of the use of academic debates as a teaching methodology | Full-length Article |
| 400 | 27 | French & Koeberl (2010) | The convincing identification of terrestrial meteorite impact structures: What works, what doesn't, and why | Review Article |
| 66 | 26 | Doody & Condon (2012) | Increasing student involvement and learning through using debate as an assessment | Full-length Article |
| 64 | 25 | Ersoy (2010) | Social studies teacher candidates' views on the controversial issues incorporated into their courses in Turkey | Full-length Article |
| 91 | 24 | Nam (2014) | The effects of trust and constructive controversy on student achievement and attitude in online cooperative learning environments | Full-length Article |
| 1292 | 23 | Kalter (2003) | Teratology in the 20th century Environmental causes of congenital malformations in humans and how they were established | Full-length Article |
| 1708 | 23 | Millan (2006) | Multi-target strategies for the improved treatment of depressive states: Conceptual foundations and neuronal substrates, drug discovery and therapeutic application | Review Article |
| 50 | 21 | Bruschke & Divine (2017) | Debunking Nixon's radio victory in the 1960 election: Re-analyzing the historical record and considering currently unexamined polling data | Full-length Article |
| 107 | 21 | Stepanova & Bruckmeier (2013) | The relevance of environmental conflict research for coastal management. A review of concepts, approaches and methods with a focus on Europe | Review Article |
| 74 | 20 | Colston & Vadjunec (2015) | A critical political ecology of consensus: On "Teaching Both Sides" of climate change controversies | Full-length Article |

*__Table SI 7.__ Publications that were most cited in the context of disagreement.*

| Total citances | Disagreement citances | Publication | Title | Document type |
|---|---|---|---|---|
| 389 | 99 | Munro (2003) | Lipid Rafts Elusive or Illusive? | Review Article |
| 477 | 99 | Murphy et al. (2009) | The impact of global signal regression on resting state correlations: Are anti-correlated networks introduced? | Full-length Article |
| 1753 | 83 | Zhao, Sun, Wilde, & Sanzhong (2005) | Late Archean to Paleoproterozoic evolution of the North China Craton: key issues revisited | Full-length Article |
| 1344 | 69 | Li et al. (2008) | Assembly, configuration, and break-up history of Rodinia: A synthesis | Full-length Article |
| 663 | 65 | Kusky & Li (2003) | Paleoproterozoic tectonic evolution of the North China Craton | Full-length Article |
| 1377 | 64 | Zhao, Wilde, Cawood, & Sun (2001) | Archean blocks and their boundaries in the North China Craton: lithological, geochemical, structural and P–T path constraints and tectonic evolution | Full-length Article |
| 972 | 51 | Zhai & Santosh (2011) | The early Precambrian odyssey of the North China Craton: A synoptic overview | Full-length Article |
| 50 | 45 | Debat et al. (2003) | A new metamorphic constraint for the Eburnean orogeny from Paleoproterozoic formations of the Man shield (Aribinda and Tampelga countries, Burkina Faso) | Full-length Article |
| 448 | 45 | Wilde, Zhao, & Sun (2002) | Development of the North China Craton During the Late Archaean and its Final Amalgamation at 1.8 Ga: Some Speculations on its Position Within a Global Palaeoproterozoic Supercontinent | Full-length Article |
| 247 | 43 | Kusky (2011) | Geophysical and geological tests of tectonic models of the North China Craton | Full-length Article |